# Distributing entanglement between distant semiconductor qubit registers using a shared-control shuttling link

Zarije Ademi,[1,*] Marion Bassi,[1,*] Cécile X. Yu,[1] Stefan D. Oosterhout,[2] Yuta Matsumoto,[1] Sander L. de Snoo,[1] Amir Sammak,[2] Lieven M. K. Vandersypen,[1] Giordano Scappucci,[1] Corentin Déprez,[1,†] and Menno Veldhorst[1,†]

[1] *QuTech and Kavli Institute of Nanoscience, Delft University of Technology, PO Box 5046, 2600 GA Delft, The Netherlands*

[2] *QuTech and Netherlands Organisation for Applied Scientific Research (TNO), Delft, The Netherlands*

Semiconductor quantum processors have potential to scale to modular quantum computers, in which qubit registers are coupled by quantum links, enabling high connectivity and space for control circuitry [1, 2]. Individual spin-qubit registers have progressed to two-dimensional systems and execution of small quantum algorithms [3–9]. Separately, high-fidelity spin shuttling has been demonstrated in linear channels defined by individual gate electrodes [10–12]. Here, we realize the first shared-control shuttling link integrated between distant qubit registers to demonstrate quantum entanglement in a basic modular quantum processor based on hole spin qubits in germanium. We develop a protocol to compensate for spin-orbit-induced rotations during qubit transfer, allowing for shuttling between qubit registers separated by more than one micrometer in approximately a hundred nanoseconds. Combining local qubit operation with coherent shuttling, we generate Bell states formed by spins residing in separate registers. Characterizing them using quantum state tomography, we demonstrate entanglement between spin qubits in distant registers.

## INTRODUCTION

Spin qubits in semiconductor quantum dots [1] hold promise for building large-scale quantum computers as they benefit from a small footprint and compatibility with semiconductor manufacturing techniques [13–15]. In the last decade, high-fidelity operations have been demonstrated in linear systems [3–6] and small two-dimensional qubit arrays [7–9]. However, scaling qubit registers further in two dimensions is non-trivial due to challenges in the interconnection [16, 17]. To overcome this bottleneck, a network approach has been proposed, where individual qubit registers are interconnected using quantum links to provide space for local electronics and wiring [2, 18].

The prospect of on-chip quantum networks has resulted in a variety of proposals to construct quantum links. Experimentally, coupling to photons in superconducting resonators enabled coherent interaction between distant qubits [19], while surface acoustic waves in gallium arsenide allowed to coherently displace electron spins [20]. Shuttling, where spin qubits are displaced using gate voltage pulses, has gained attraction due to its compatibility with semiconductor quantum dot devices. Crucially, spin qubits can be shuttled with high fidelity [10–12], enabling the implementation of quantum links. Shuttling can furthermore be used for the construction of single-qubit [10] and two-qubit gates [21, 22]. However, to make shuttling an effective approach for scaling quantum networks, it is crucial to advance beyond individually wiring each gate of the shuttling channel. Moreover, quantum networks will require interfacing of qubit registers and quantum links, while their co-integration comes with further challenges in both the fabrication and operation.

Here, we show that spin qubit shuttling can be executed with shared control and co-integrated to provide a quantum link between distant qubit registers. Our rudimentary quantum network is based on hole spins in germanium, where the strong spin-orbit interaction leads to qubit dynamics during transfers between registers, which can be used for fast qubit control. We investigate both the diabatic and adiabatic spin-shuttling regimes. In the former, we develop a protocol that compensates spin-orbit-driven qubit rotations. Combining local control with shuttling, we demonstrate entanglement between spin qubits in separate registers. The resulting Bell-state fidelities, quantified using tomography, reach 61%-67% and up to 82%-92%, after correcting for errors in state preparation, measurement and recovery operation, setting a benchmark for remotely entangled semiconductor qubits.

## QUBIT REGISTERS

Our minimal quantum network consists of two registers interconnected by a quantum link, see Fig. 1.a. The device is fabricated using an overlapping gate technique [23, 24] on a strained germanium heterostructure [25] (see Supplementary Section 1 A for details). Both quantum registers are composed of three quantum dots denoted D*i*. The registers are coupled by a shuttling lane, where spins can be transferred using a traveling-wave potential generated by phase-shifted sinusoidal voltage pulses applied on successive series of clavier gates (Cl*i*) [11, 18, 26–30]. In our device, the clavier gates are galvanically connected on the chip, such that the length

* These authors contributed equally
† These authors jointly supervised this work

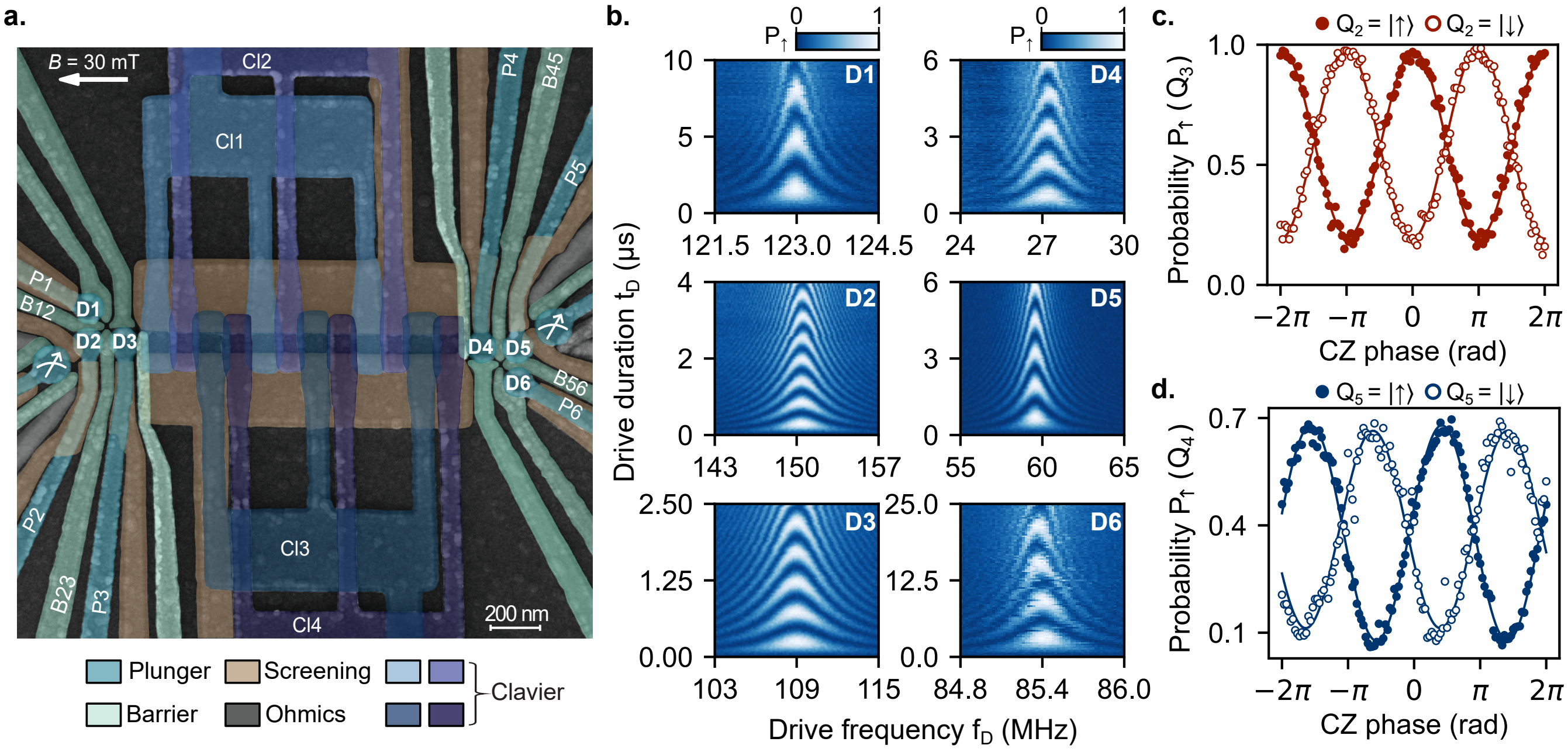

Figure 1. **Shared-control shuttling link between spin-qubit registers in germanium.** **a.** False-colored scanning electron micrograph of a modular quantum device, nominally identical to the device studied. The device consists of a 1.25 μm long shuttling lane that connects two qubit registers. Each register is made of three quantum dots and has a dedicated charge sensor. Quantum dots, labeled D$i$, are formed underneath the plunger gates P$i$ while the barrier gates, denoted B$ij$, control the tunnel coupling between the quantum dots D$i$ and D$j$. The shuttling lane is composed of two wide screening gates forming a one-dimensional channel between the qubit registers. The shuttling phase-shifted sine waves are applied to the clavier gates Cl$i$. **b.** Rabi chevrons measured for qubits $Q_i$ in the six quantum dots D$i$. Evolution of spin-up probability $P_\uparrow$ as function of the duration and frequency of the driving pulse. **c.** Controlled-$Z$ gate performed between qubits $Q_2$ and $Q_3$ located in D2 and D3 respectively. After initialization, $Q_3$ is prepared in a superposition state and accumulates a conditional phase of 0 or $\pi$ resp.,depending on the $Q_2$ state, $|\downarrow\rangle$ or $|\uparrow\rangle$ resp., probed by a Ramsey-type of experiment. Here is shown the up-state probability of $Q_3$ as a function of a phase applied before projecting the qubit back and reading its state. Depending on the $Q_2$ state, the target qubit $Q_3$ phase is shifted by about $\pi$, demonstrating a successful $CZ$ gate calibration. **d.** Controlled-$Z$ gate performed in the second register between $Q_4$ and $Q_5$ respectively located in dots D4 and D5. Experiment carried out here is similar to the one in (c): $Q_4$ experiences a phase shift of $\pi$ depending on $Q_5$ state.

of the shuttling lane can be scaled independently of the number of control terminals.

Each dot is filled with an odd number of holes (either 1 or 3), and the spins are initialized pairwise in their $|\downarrow\downarrow\rangle$-state by pulsing adiabatically from the (1,3) to the (2,2) or (0,4) charge state and back (further details in Supplementary Section 1 C 3). We then operate the qubits by Electric Dipole Spin Resonance (EDSR), giving rise to Rabi oscillations as shown in Fig. 1.b (see Supplementary Table 1 and Supplementary Table 2 for further details on the qubit metrics). Spin qubits in neighboring quantum dots D2/D3 (D4/D5) can be coupled by dynamically tuning their exchange interaction via the barrier gate B23 (B45), therefore allowing for controlled-$Z$ gate operations as shown in Fig. 1.c and Fig. 1.d. Here, the main focus is laid on creating entanglement between spin qubits residing in different registers, where quantum dots D2, D3, and D4 play a pivotal role. In the following experiments, the qubit $Q_A$ remains stationary in D2 while the qubit $Q_B$ serves as a mobile qubit that will be shuttled between D3 and D4.

## COHERENT SPIN SHUTTING BETWEEN REGISTERS

To transfer $Q_B$ from the left register to the right register, we proceed in three main steps : (i) the spin is initially loaded from D3 into the shuttling lane, then (ii) it is conveyed within the moving electrostatic potential created by the clavier gates and (iii) finally it is unloaded into D4, such that it arrives in the second qubit register. Quantum dots D3 and D4 are tuned to be highly tunnel-coupled to the first and last sites of the conveyor lane, as evidenced by the stability diagrams in Figs. 2.e and 2.f. This configuration enables loading and unloading the conveyor lane through changes of the gate voltages controlling the energy detuning between the quantum dots as illustrated in Fig. 2.a.

We study the spin coherence during the (un)loading process by conducting the shuttling experiments sketched in Fig. 2.c-d. The spin is prepared in a superposition state in D3 (D4), before being transferred to one end of the conveyor channel using a detuning pulse. After a wait time $t$, the spin is returned to its initial position using a reversed detuning pulse and a second $\pi/2$ pulse is applied. Fig. 2.g-h evidence the resulting up-state probability ($\mathrm{P}_\uparrow$) as a function of the wait time for different amplitudes of the detuning pulse. Once in the conveyor lane, the spin qubit accumulates a phase due to the difference between its initial Larmor frequency in D3 (D4) and the local Larmor frequency in the conveyor lane [31, 32]. The observation of time-resolved oscillations resulting from such phase accumulation confirms that the qubit remains coherent throughout the (un)loading process.

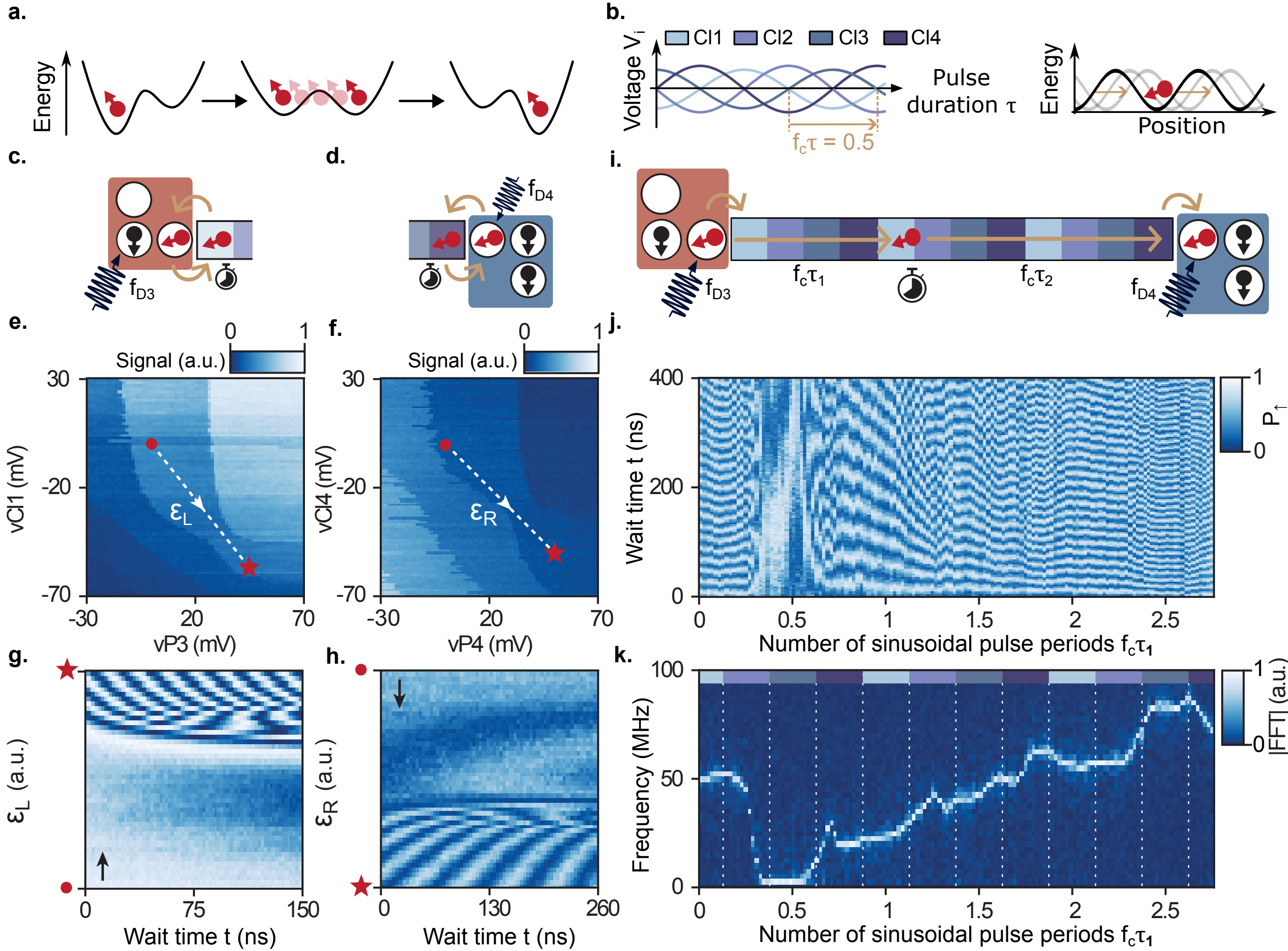

Figure 2. **Integration and characterization of the shuttling lane. a.** Schematic showing bucket-brigade shuttling for conveyor lane loading and unloading. Changes in detuning energy between two quantum dots allow to transfer the spin. **b.** For conveyor shuttling, phase-shifted sinusoidal pulses are applied to each clavier gate in order to create a moving electrostatic potential. **c.-d.** Schematics of the free-evolution experiments used to investigate the qubit coherence during (un)loading protocol into the shuttling lane. **e.-f.** Stability diagram of D3 (D4) and the first (last) site in the shuttling lane, located underneath the clavier gate Cl1 (Cl4). The spin (un)loading is performed by pulsing along the detuning axis ($\epsilon_R$) $\epsilon_L$ marked by the white dashed line. **g.-h.** Corresponding coherent time-resolved oscillations in up-state probability ($\mathrm{P}_\uparrow$) recorded as function of the amplitude of the detuning voltage pulse. **i.** Schematic of the free-evolution experiment used to investigate the coherence during the qubit transfer between the registers. **j.** Corresponding time-resolved oscillations of $\mathrm{P}_\uparrow$ recorded as a function of the number of periods $f_c\tau_1$ in the first conveyor pulse. Measurements were performed at a shuttling frequency $f_c = 0.5\,\mathrm{MHz}$ corresponding to the adiabatic regime. **k.** Fast Fourier Transform (FFT) of the oscillations presented in (j), highlighting the local Larmor evolution, with respect to the initial Larmor frequency $f_L^{D3} = 110$ MHz, and the displacement of the spin qubit as a function of $\tau_1$. The blue rectangles indicate the clavier gate beneath which we expect the minima of the traveling-wave potential to be located depending on $f_c\tau_1$. The dashed lines mark the transitions between clavier gates.

We now focus on spin transfers between the registers. The moving electrostatic potential is generated by applying voltage pulses $V_i(\tau)$ to the clavier gate Cl$i$, given by:

$$V_i(\tau) = -A_i \sin\left(i\frac{\pi}{2} - 2\pi f_c \tau\right), \tag{1}$$

where $f_c$ and $A_i$ are respectively the frequency and the amplitude of the pulses applied (Fig. 2.b). Different regimes occur depending on the chosen frequency and amplitude (see Supplementary Section 3). We find that when the pulse amplitudes are sufficiently large, spins can be transferred from one register to another and their dynamics mainly depend on the frequency $f_c$ that controls the shuttling speed.

When the shuttling speed is low, in our experiments when $f_c \lesssim 1$ MHz, the shuttling is adiabatic with respect to the charge and spin degree of freedom. In this regime, we can verify that the phase coherence is preserved by performing another free-evolution experiment illustrated in Fig. 2i. After loading a superposition state from D3, the spin is shuttled for a duration $\tau_1$ by applying the sine pulses on the clavier gates. It reaches an intermediate position within the conveyor lane, where it is allowed to precess freely for a time $t$ by holding the voltages applied on the clavier gates. Then the qubit is shuttled for a complementary duration $\tau_2$ such that $f_c(\tau_1 + \tau_2) = 2.75$. At this point, we expect the spin to be under the rightmost electrode of the conveyor channel. It can thus be unloaded into D4, where a second $\pi/2$ pulse is applied before determining the qubit state. The resulting $P_\uparrow$ evolution measured as a function of number of periods $f_c\tau_1$ in the first shuttling pulse and of the precession duration $t$ is presented in Figure 2.j. The presence of oscillations for any values of $f_c\tau_1$ confirms that spin coherence is preserved during transfer between registers.

Large variations of the oscillation frequency $f_{osc}$ are observed along the conveyor lane in Figure 2.j, which can be further highlighted by computing the corresponding Fast Fourier Transform (FFT) (Fig. 2.k). The oscillation frequency is given by the difference between $f_L^{D3}$ the Larmor frequency in D3 and the frequency $f_L(\tau_1)$ at the location in the conveyor where the spin idles, so that $f_{osc}(\tau_1) = |f_L^{D3} - f_L(\tau_1)|$ (see Supplementary Section 6 A). The latter varies due to the strong dependence of the $g$-tensor on the local confining potential, a hallmark of the spin-orbit interaction in hole spin qubits. The frequency evolution therefore evidences that the qubit experiences different strain and electrostatic environments as a function of $f_c\tau_1$ as expected for a shuttled spin.

## SHUTTLING-INDUCED ROTATIONS DUE TO SPIN-ORBIT INTERACTION

Another consequence of the strong spin-orbit interaction in germanium is the presence of sizable variations of the spin quantization axis between quantum dots [33] which can lead to qubit rotations occurring during fast shuttles. Although these rotations can be leveraged for high-fidelity qubit operations [10], they must be characterized and compensated for the diabatic shuttling to be useful for quantum information transfer.

Since we have universal qubit control in each register, we explore the use of tailored pulses to undo the unitary operation $R_{shut}$, experienced by the qubit during shuttling. We investigate this through the experiments illustrated in Fig. 3.a in which $Q_B$ is shuttled back-and-forth between D3 and D4 at a high conveyor frequency $f_c = 50$ MHz. During each round trip, the qubit undergoes a unitary operation $R_{shut}$ due to local variations in the quantization axes and the Larmor frequencies. After the shuttling, we compensate $R_{shut}$ with a recovery operation $R_{rec}$ implemented via an adapted EDSR pulse applied in D3.

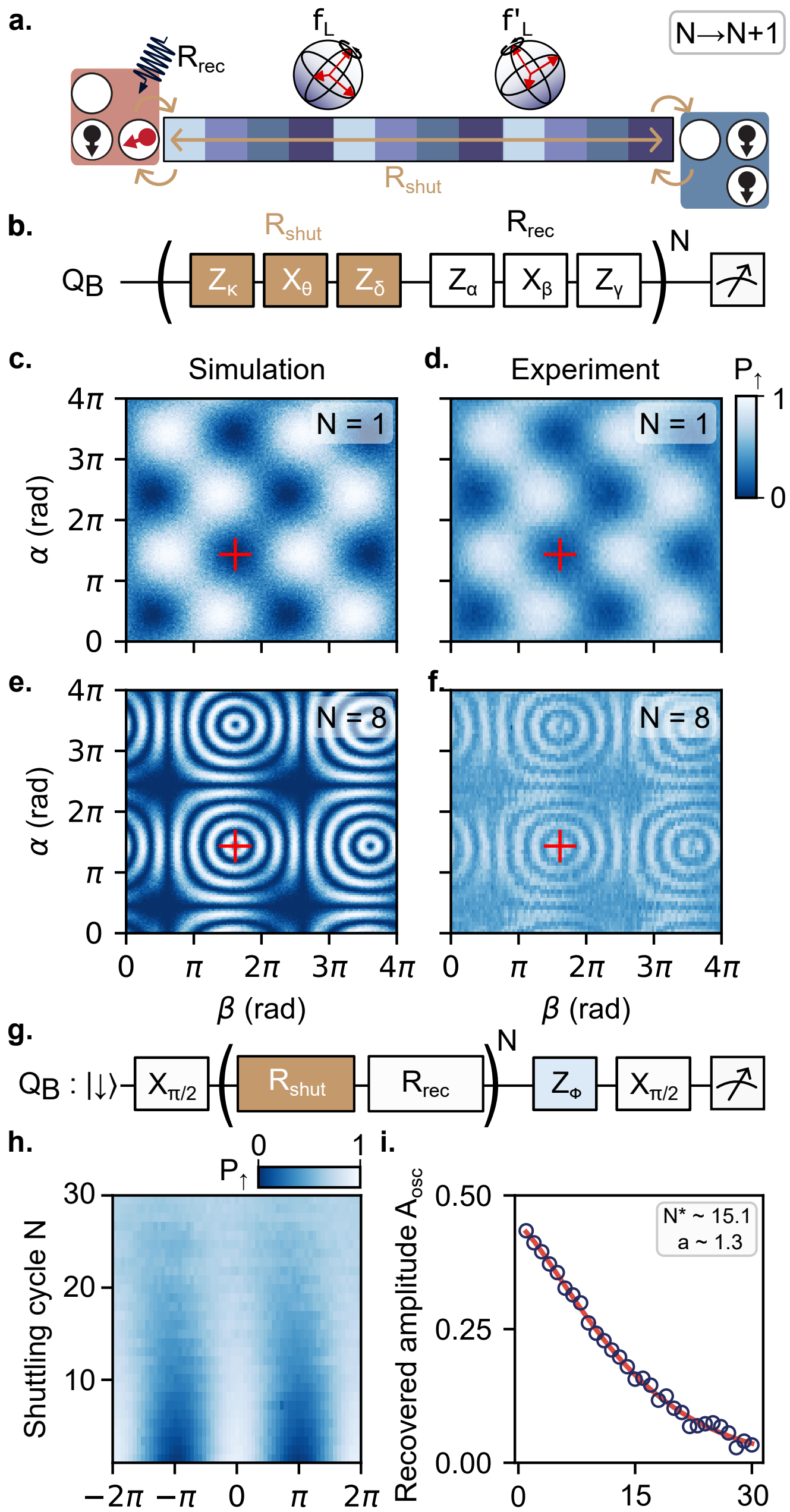

Figure 3. **Diabatic spin shuttling and compensation of spin-orbit-induced rotations. a.** Schematic of the experiments used to investigate diabatic shuttling and the compensation of spin-orbit-induced rotations. A qubit is diabatically shuttled back-and-forth between D3 and D4. Meanwhile, it experiences a rotation $R_{shut}$ due to local variations of the quantization axes and Larmor frequencies. $R_{shut}$ is compensated by a recovery gate $R_{rec}$ applied in D3. The overall cycle formed by the back-and-forth shuttling process and the recovery operation is repeated $N$ times. **b.** Corresponding quantum circuit and decomposition of $R_{shut}$ and $R_{rec}$. **c.-f.** Calibration of $\alpha$ and $\beta$. The right column presents the spin up-state probability $P_\uparrow$ measured for $N = 1$ and $N = 8$ cycles. The left column shows noise-free simulations of the expected experimental results. The $\kappa$, $\theta$ and $\delta$ angles used in the simulations are extracted from the calibration procedure. The red cross marks the combination of $\alpha$ and $\beta$ angles conserving the initial $|\downarrow\rangle$ state. **g.** Circuit used to characterize the coherence of qubit $Q_B$ after $N$ shuttling cycles. **h.** Oscillations in up-state probability $P_\uparrow$ measured in such experiment as a function of $\Phi$ for increasing $N$. **i.** Recovered amplitude $A_{osc}$ of the oscillations presented in (h), fitted with an exponential decay $A_{osc}(N) = A_0 \exp(-(N/N^*)^a) + A_{offset}$ (solid line).

Since the shuttling and recovery gates are unitary operations, they can be decomposed as a combination of $X$ and $Z$ rotations [34, 35], reading as $R_{shut} = Z_\delta X_\theta Z_\kappa$ and $R_{rec} = Z_\gamma X_\beta Z_\alpha$ (Figure 3.b). While the $(\kappa, \theta, \delta)$ angles are a priori unknown and set by the shuttling sequence, the recovery-gate angles $(\alpha, \beta, \gamma)$ can be tuned from 0 to $2\pi$ by changing respectively the initial phase, the duration of the compensation pulses and the phase of all subsequent EDSR pulses. Thus, one can always find a set of recovery-gate angles $(\alpha, \beta, \gamma)$ such that $R_{rec}R_{shut}$ gives an identity operation.

We can calibrate these angles using a two-step protocol, that also applies for one-way shuttles. First, we initialize qubit $Q_B$ in its $|\downarrow\rangle$ state, shuttle it back-and-forth and apply a recovery pulse. Since the spin is in its ground state, the $Z$ rotations with angles $\kappa$ and $\gamma$ have no influence and can be disregarded. The two remaining angles $\alpha$ and $\beta$ are varied to identify the values that maximize the return probability $P_\downarrow = 1 - P_\uparrow$. Simulations of such a quantum circuit and the corresponding measurements are displayed in Fig. 3.c-d where the optimal angles are marked with the red cross. Once $\alpha$ and $\beta$ are calibrated, we determine the angle $\gamma$ by maximizing $P_\uparrow$ in a Ramsey-like experiment where the shuttling and recovery operations are inserted between two $X_{\pi/2}$ gates.

We validate the recovery-gate calibration by repeating the shuttling and recovery sequence $N$ times to amplify any calibration error. Fig. 3.e-f show the simulated and measured up-state probabilities for $N = 8$ cycles. Due to error amplification, the dependence of $P_\uparrow$ on the angles $\alpha$ and $\beta$ now exhibits concentric ring-shaped interference patterns. The center of these patterns highlights the values of $\alpha$ and $\beta$ that render the spin state invariant under repeated shuttling. They align with the calibrated values of $\alpha$ and $\beta$ (red cross) confirming the successful calibration of the recovery gate. We note that for $N > 1$, additional waiting times in D3 are required after the recovery gates to realign the qubit's rotating frame in the laboratory frame before performing the next shuttle (see Supplementary Section 7 B). Overall, the remarkable agreement between the simulations and the experiments shows that shuttling-induced rotations are well-described by the aforementioned gate decomposition and that their interplay with rotations induced by resonant pulses is understood.

Importantly, combining diabatic shuttling with recovery operation allows to drastically reduce the transfer time between registers, from 6.7 µs in Fig. 2.j-k to 110 ns in Fig. 3 and thus to limit the impact of dephasing during the transfers.

## EVALUATION OF THE TRANSFER PERFORMANCE

To quantify the fidelity of the qubit transfer, we perform the experiments sketched in Fig. 3.g in which the shuttled qubit $Q_B$ is prepared in a superposition state and is subjected to $N$ repeated shuttling cycles. After the final recovery operation, we apply a phase gate $Z_\Phi$ together with a $X_{\pi/2}$ gate to probe the remaining coherence. The measured up-state probability as a function of angle $\Phi$ and of the number of cycles $N$ is shown in Figure 3.h. As expected for a successful calibration, $P_\uparrow$ reaches a maximum at $\Phi = 0$ rad indicating that the combined operation $R_{rec}R_{shut}$ closely approximates the identity operation. The oscillations fade out for increasing $N$, reflecting the qubit decoherence experienced during the repeated transfers: their amplitude $A_{osc}$ decays exponentially with $N$ as shown in Fig. 3.i.

The fitted decay constant $N^* = 15.08 \pm 0.43$ can be converted into an effective transfer distance $d^* = 2N d_{D3/D4} \simeq 43$ µm assuming that the spins travel between the centers of D3 and D4 which are designed to be spaced by $d_{D3/D4} = 1.43$ µm. While this effective distance is significantly below those reported in most recent experiments performed in silicon-based devices ($d^* \simeq 2.5$ mm) [11], we emphasize that here we focus on the transfer between qubit registers, rather than the shuttling of a spin within a shuttling channel. The total number of coherent back-and-forth transfers achievable here is affected by additional sources of errors compared to prior works. In particular, it is degraded by (i) errors in the calibration of the recovery gate, (ii) imperfections during the loading and unloading steps and (iii) errors arising from the abrupt application of voltage pulses to the clavier gates. For an increasing number of cycles $N$, these errors get progressively amplified. Furthermore, we emphasize, that in this case, the qubits are shuttled over a real distance of 1.43 µm, few times longer than in previous works [11, 29, 30]. Thus we are more likely to encounter weak points in this device. This effect may be mitigated by optimizing the shape of the conveyor pulse as demonstrated in refs. [11, 36]. Such an optimization may also be used to further increase the shuttling frequency and to reduce dephasing upon shuttling.

## REMOTE ENTANGLEMENT BETWEEN DISTANT QUBIT REGISTERS

In modular quantum architectures comprising interconnected qubit registers, the ability to distribute entangled qubits between registers is a key requirement. In this section, we demonstrate the generation of entanglement between spins in the two spatially separated registers based on local entangling gates and shuttling. For this purpose, we implement the circuit presented in Figure 4.a. All qubits present in the left and right registers (i.e. in quantum dots D1, D2, D3, D5, D6) are initialized in their ground state. A Bell state, denoted $|\Phi^{\pm}\rangle$ or $|\Psi^{\pm}\rangle$, is then prepared by combining single- and two-qubit gates on qubit $Q_A$ in dot D2 and qubit $Q_B$ in dot D3. Qubit $Q_B$ is then diabatically shuttled within the conveyor lane into dot D4 prior to the recovery gate. Additional phase corrections $Z_\theta$ and $Z_\eta$ are applied on both qubits to ensure synchronization of the qubit rotating frames (see Supplementary Section 6). $Z_\theta$ also compensates the phase shifts acquired by $Q_A$ while $Q_B$ is shuttled. At this point, the joint state of the spins in D2 and D4 should ideally be the initialized Bell state. We monitor the remaining entanglement by performing Quantum State Tomography (QST) measurements and by reconstructing the two-qubit density matrix.

Fig. 4.b-e illustrate the measured density matrices ($\rho$) for each prepared Bell states and the corresponding fidelity $F$ estimated by the comparison with the ideal matrices ($\sigma$) as $F(\rho,\sigma) = \left(\mathrm{Tr}\sqrt{\sqrt{\sigma}\rho\sqrt{\sigma}}\right)^2$. The mean bare fidelities reaches on average about 64.0% and up to 87.1% after compensation for state preparation, measurement and recovery gate errors (see Supplementary Section 8). These results prove that semiconductor spin qubits in distant registers can be entangled.

## CONCLUSION

In this work, we have demonstrated the implementation of a minimal on-chip quantum network based on germanium spin-qubit registers in which the quantum link is implemented via spin shuttling. We have shown that coherent shuttling can be realized with shared control, which may become essential in the scale-up toward modular quantum processors. We proposed and demonstrated an experimental protocol to compensate for spin-orbit-induced rotations using resonant pulses, therefore enabling for spin transfer over about 1.5 μm on timescales of about 100 ns. Building on these results, we demonstrated quantum information distribution by generating remote entanglement after shuttling one of the entangled qubits. Research on spin qubits has seen exciting progress in individual elements, including control and readout of qubit registers registers [3–10, 37] and shuttling of single spins [10, 11, 21, 29, 32]. By integrating these elements into a rudimentary quantum network, this work opens new prospects for modular quantum computing using semiconductor spins.

## ACKNOWLEDGMENTS

We thank the members of Veldhorst, Vandersypen and Scappucci groups for fruitful discussions and help provided all along the project. We thank M. Rimbach-Russ and S. Bosco for insightful discussions. This research was supported by the European Union through the Horizon 2020 research and innovation program under the Grant Agreement No. 951852 (QLSI) and the Horizon Europe Framework Program under grant agreement No. 101069515 (IGNITE). M. V. acknowledges support from the NWO through the National Growth Fund program Quantum Delta NL (grant NGF.1582.22.001). This research was sponsored in part by the Army Research Office (ARO) under Award No. W911NF-23-1-0110 and by The Netherlands Ministry of Defense under Awards No. QuBits R23/009. The views, conclusions, and recommendations contained in this document are those of the authors and are not necessarily endorsed nor should they be interpreted as representing the official policies, either expressed or implied, of the Army Research Office (ARO) or the U.S. Government, or The Netherlands Ministry of Defense. The U.S. Government and The Netherlands Ministry of Defense are authorized to reproduce and distribute reprints for Government purposes notwithstanding any copyright notation herein.

## AUTHOR CONTRIBUTIONS

Z. A., M. B. and C. D. performed the measurements and analyzed the data. C. X. Y. and C. D. designed the device and built the experimental setup. A. S. and G. S. supplied the heterostructure and S. D. O. fabricated the sample. S. L. d. S. developed the software with contributions from Y. M., supervised by L. M. K. V.. Z. A., M. B., C. D. and M. V. wrote the manuscript with inputs from all authors. C. D. and M. V. conceived and jointly supervised the project.

## COMPETING INTERESTS

M. V., G. S. and L. M. K. V. are founding advisors of Groove Quantum BV and declare equity interests. The remaining authors declare that they have no competing interests.

## CORRESPONDING AUTHORS

Correspondence should be sent to Corentin Déprez (corentin.deprez@neel.cnrs.fr) or Menno Velhorst (M.Veldhorst@tudelft.nl).

## DATA AVAILABILITY

All data underlying this study are available in an online repository at 10.4121/1a4557e5-d9cd-4727-a8ab-ed9df7c790f6.

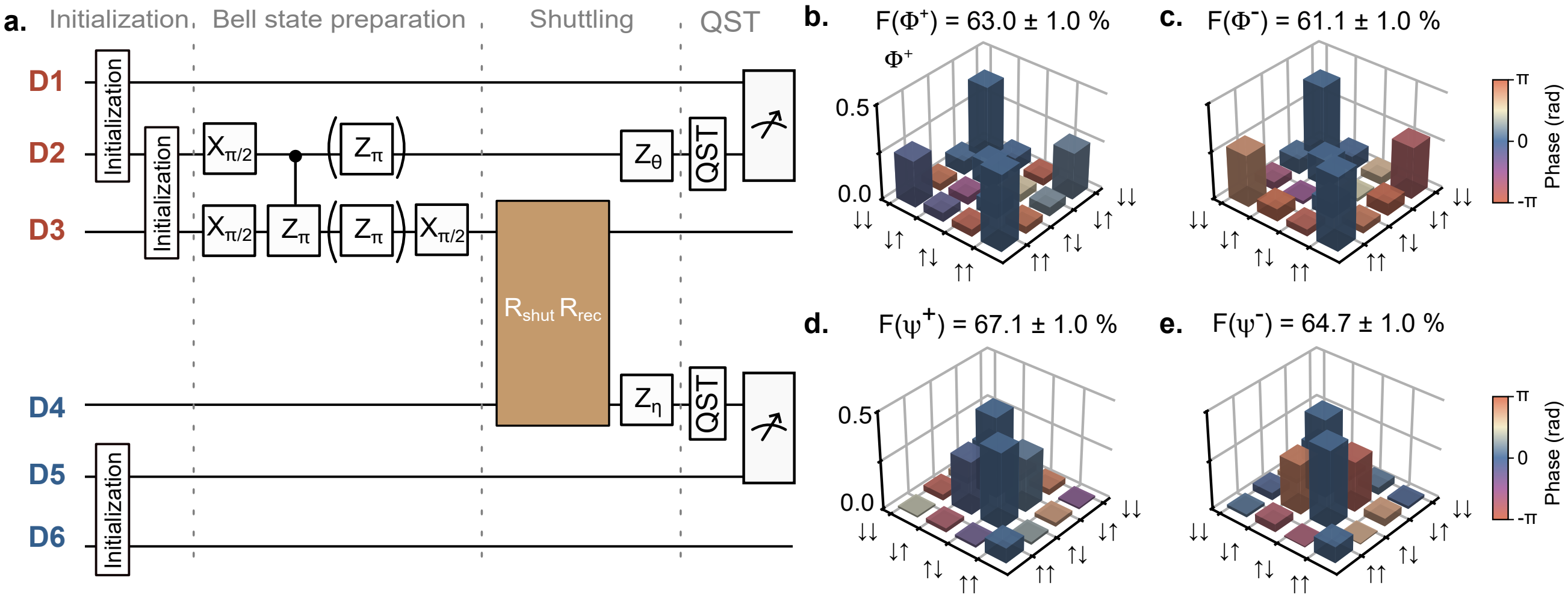

Figure 4. **Entanglement between spins in distant qubit registers enabled by shuttling. a.** Quantum circuit used to assess the presence of remote entanglement while shuttling the spin $Q_B$ from D3 to D4. After the initialization stage and the preparation of the entangled state, the qubit $Q_B$, initially located in D3, is shuttled diabatically to D4. To account for shuttling-induced rotations, synchronization of rotating frames and single-qubit phase shifts accumulated during the state preparation and the transfer, the recovery gate and some virtual $Z$ rotations are applied before QST measurements (see Supplementary Materials). **b.-e.** Reconstructed density matrices of the four Bell states and corresponding Bell state fidelities. Fidelity errors are calculated by bootstrapping analysis (see Supplementary Section 8) and give to the 95% confidence intervals.

# Supplementary Information for: Distributing entanglement between distant semiconductor qubit registers using a shared-control shuttling link

Zarije Ademi,[1,*] Marion Bassi,[1,*] Cécile X. Yu,[1] Stefan D. Oosterhout,[2] Yuta Matsumoto,[1] Sander L. de Snoo,[1] Amir Sammak,[2] Lieven M. K. Vandersypen,[1] Giordano Scappucci,[1] Corentin Déprez,[1,†] and Menno Veldhorst[1,†]

[1] *QuTech and Kavli Institute of Nanoscience, Delft University of Technology, PO Box 5046, 2600 GA Delft, The Netherlands*

[2] *QuTech and Netherlands Organisation for Applied Scientific Research (TNO), Delft, The Netherlands*

This Supplementary Information includes :

- Supplementary Sections 1-8
- Supplementary Figures 1-14
- Supplementary Tables 1-3
- Supplementary References 1-30

* These authors contributed equally

† These authors jointly supervised this work

## Supplementary Section 1. METHODS

### A. Materials and device fabrication

The device is fabricated on a Ge/SiGe heterostructure grown by reduced pressure chemical vapor deposition on a silicon wafer. The strained germanium quantum well is buried below a 55 nm thick SiGe layer capped with a sacrificial Si layer [1]. The device is fabricated by successive steps involving electron beam lithography, metal evaporation, lift-off and deposition of alumina dielectric layers. The ohmic contacts are first made by depositing a 30 nm thick platinum film after etching of the sacrificial Si cap. Subsequently, a 7 nm thick $Al_2O_3$ layer is deposited by atomic layer deposition at 300 °C. The high temperature used for the oxide deposition allows the diffusion of the platinum into the SiGe barrier and to contact the Ge quantum well underneath. The gate stack is then defined by successive depositions of four Ti/Pd bi-layers (with respective thicknesses of 3/17, 3/27, 3/37 and 3/37 nm) each separated by 5 nm thick alumina layers. The screening gates are first fabricated, followed by the barrier gates together with the clavier gates Cl1 and Cl3. Subsequently, the plunger gates and the clavier gates Cl2 and Cl4 are made. The last step consists in depositing the barrier gates that separate the conveyor lane from the qubit registers.

### B. Measurement setup

The measurements are performed in a dilution fridge with a base temperature of about 15 mK. It is equipped with a single-axis superconducting coil used to generate the external magnetic field of 30 mT. The electrostatic landscape in the device is shaped by DC voltages applied to the gate electrodes using 18-bit digital-to-analog converters (D5a modules of the SPI rack). The DC lines are filtered inside the dilution refrigerator, at the lowest temperature stage, using two-stage low-pass RC filters and a copper powder filter.

In addition, AC voltage pulses are applied for qubit operation (EDSR driving, initialization and readout) and shuttling. These pulses are generated by QBlox QCM modules whose time resolution are enhanced to about 50 ps using the method described in refs [2, 3]. The latter pulses are attenuated both at room temperature and within the dilution fridge with a total attenuation ranging typically from 18 to 26 dB and up to 52 dB in the case of the conveyor screening gates. Finally, these AC voltage pulses are combined with the DC signals using on-PCB bias-tees. All pulse sequences are programmed, compiled and generated on the hardware using “core-tools” (https://gitlab.tudelft.nl/qutech-qdlabs/core_tools), “pulse_lib” (https://gitlab.tudelft.nl/qutech-qdlabs/pulse_lib) and “qconstruct” python libraries.

As further discussed in Supplementary Section 1 C 3, spin-qubit readout relies on spin-to-charge conversion and charge sensing, the latter being performed by radio-frequency reflectometry. Each qubit register is accompanied by a nearby charge sensor, consisting of a Single Hole Transistor (SHT), with one of its ohmic contacts connected to an on-PCB NbTi inductor. Together, the inductor and the charge sensor in the red register (resp. the blue register) form a tank circuit whose resonance frequency is about 93.7 MHz (resp. 109.5 MHz). The tank circuits are all connected to a common readout line on the PCB and frequency multiplexing is used for the readout.

The probe signal has an amplitude of typically 100 mV and is generated by a QBlox QRM module. It is attenuated at room temperature (50 dB) and throughout the dilution refrigerator (23 dB) before being redirected to the device by a directional coupler (Mini Circuits ZEDC-15-2B) at the mixing chamber stage. Once reflected, the signal is amplified at the 4 K stage (by a Cosmic Microwave Technology CITLF3 amplifier with a typical gain of 30 dB) and at room temperature (by a M2j module of the SPI rack with a variable gain between 35 and 70 dB) before the demodulation.

### C. Device operation

#### 1. *Tuning of the electrostatic potential landscape*

Initially, the DC voltages are tuned such that quantum dots are formed under each plunger gates P1-P6 and at both ends of the conveyor channel, i.e. under the leftmost electrode of Cl1 and the rightmost electrode of Cl4. Quantum dots D2 and D5 are tuned to host three holes whereas D1, D3, D4 and D6 are tuned either to be depleted or to host one hole depending on the experiment. The quantum dots at the edges of the conveyor channel are always depleted initially. During the tune-up, the clavier-gate voltages applied to the electrodes in the same layer (Cl1/Cl3 and Cl2/Cl4) are set to remain within a close range (about 150 mV difference). This ensures that a traveling-wave potential can be formed by applying the conveyor pulses. Likewise, the conveyor screening-gate voltages are set close to their turn-on values to facilitate the formation of the moving potential.

### 2. Gate voltage virtualization

After the initial DC tuning, we define virtual plunger-gate voltages vP$i$, virtual barrier-gate voltages vB$ij$ and virtual clavier-gate voltages vCl$i$ by making linear combinations of the real plunger- P$i$, barrier- B$ij$ and clavier-gate voltages Cl$i$. This allows to compensate for gate-to-gate cross-capacitances and it provides independent control over each quantum-dot chemical potential. It also ensures that the interdot tunnel couplings can be tuned without changing the quantum-dot chemical potentials. We note that the virtual barrier-gate voltages do not allow to have fully independent control over the interdot tunnel couplings within a register. Similarly, each charge-sensor plunger gate is virtualized with respect to the neighboring gates to ensure a good charge sensitivity over a wide range of gate voltages.

### 3. Qubit operation

The spin qubits in the different quantum dots are initialized pairwise, the pairs being Q1-Q2, Q2-Q3, Q4-Q5 and Q5-Q6. The initialization protocol consists in ramping adiabatically from the (4,0) to the (3,1) charge states, or equivalently from the (2,2) to the (3,1) charge states depending on the pair, and leaves the qubit pair in the $|\downarrow\downarrow\rangle$ state. For the readout, the system is adiabatically ramped to the interdot charge transition, enabling spin-to-charge conversion via Pauli Spin Blockade (PSB) mechanism. The charge transitions are probed using RF-SHTs, as mentioned in Supplementary Section 1 B. The integration times used vary between 1 µs and 20 µs depending on the pair probed.

Single-qubit gates are performed by Electric Dipole Spin Resonance (EDSR) consisting in applying a resonant microwave tone to a gate nearby the qubit. By pulsing on the barrier gates, the exchange couplings between neighboring spin qubits can be switched on allowing for two-qubit gate operations. To enhance the controlled-$Z$ gate fidelity, the barrier-gate pulses are shaped with a Hamming windows [3].

In experiments involving shuttling in the conveyor, or more generally large-amplitude or long-duration voltage pulses, we observe large transient drops of the sensors' sensitivities that recover only after hundreds of microseconds. Consequently, in such experiments, the qubit state is measured 100 µs to 300 µs after the end of the qubit shuttling. During this time, the qubits are kept idling nearby the center of the (3,1) charge state. As the relaxation times reported in similar conditions are about milliseconds [4], these additional waiting times do not affect significantly the measured spin probabilities.

### 4. Shuttling operation

The spin transfer from the left register to the right register comprises three main steps: loading, shuttling within the conveyor lane and unloading (see main text). The first and last steps are performed by bucket-brigade shuttling. Once the spin is loaded into the conveyor, large positive voltage pulses (typically about 50 to 200 mV) are applied to close-by plunger and barrier gates in order to isolate the conveyor lane from the registers.

During the shuttling, phase-shifted sinusoidal voltage pulses $V_i(\tau)$ are applied to the clavier gate Cl$i$ given by:

$$V_i(\tau) = -A_i \sin\left(i\frac{\pi}{2} - 2\pi f_{\mathrm{c}}\tau\right) = -A_i \sin\left(\phi_i(\tau)\right), \tag{1}$$

where $A_2 = A_4 = 192$ mV and $A_1 = A_3 = 160$ mV. The ratio $A_2/A_1 = A_4/A_3 = 1.2$ is empirically chosen to compensate for the difference in lever arms between the clavier gates fabricated in different layers. This value is in line with those reported in similar experiments [5, 6].

For $f_{\mathrm{c}}\tau = (i-1)/4 + k$ with $k$ an integer, we have $\phi_i(\tau) = \pi/2 - 2k\pi$ and thus the minima of the traveling-wave potential are located under the clavier gate Cl$i$. In particular, when $f_{\mathrm{c}}\tau = k$ the potential minima are under Cl1 while for $f_{\mathrm{c}}\tau = k + 0.75$ they are under Cl4. Considering this and the design of the device, a pulse of duration $\tau_{\mathrm{shuttle}}$ satisfying $f_{\mathrm{c}}\tau_{\mathrm{shuttle}} = 2.75$ should enable to displace a qubit along the entire conveyor, from the leftmost electrode of Cl1 to the rightmost electrode of Cl4, in absence of disorder.

Equation (1) applies for a single shuttle from left to right starting with a qubit under the leftmost electrode of Cl1. When the qubit is already inside the conveyor or under another gate, an additional phase offset has to be added. This is particularly relevant in free-evolution experiments where the qubit is let to precess between two shuttling pulses. In this case, the offset should be chosen such that the initial phase of the second pulse matches the final phase of the first pulse. This ensures that the traveling-wave potential minima occur at the same locations at the end of the first shuttling pulse and at the beginning of the second one. Likewise, to enable shuttling from right to left, the sign of the $2\pi f_c \tau$ term must be changed.

In our experiments, the shuttling pulses are applied to the real clavier gates (i.e. non-virtualized ones) which causes undesired shifts of the chemical potentials of the quantum dots. This can lead to a randomization of the spin states of the static qubits used as readout ancilla and thus to degradations of the readout fidelities. In order to account for these cross-talks, calibrated sinusoidal voltage pulses are sent to the virtual plunger gates of the ancilla qubits simultaneously with the shuttling pulses.

### D. Simulations of recovery-gate calibration experiments

To simulate and reproduce the results of the experiments used to calibrate the recovery gate (Fig. 3.c-f of the main text), we use the built-in simulator of the "qconstruct" python library which is also the library used to execute quantum circuits on the real device.

To simulate a given quantum circuit, we provide it to the simulator using the same instructions than the ones used to implement it on the quantum hardware. Based on the matrix representation of the operations performed, the simulator calculates the (multi-)qubit state vector at the end of the circuit assuming that the unitary operations are ideal, that there is no decoherence and that the qubits are initialized in their $|0\rangle$ state. The simulator then reproduces the results of single-shot measurements by projecting (multi-)qubit state onto one of the basis states. The projection probabilities are determined by the squared magnitudes of the corresponding components in the state vector. Repeating the simulation typically a few hundred or a few thousand times, the probabilities of measuring the system in each basis state at the end of the circuit are finally inferred.

## Supplementary Section 2. CHARACTERIZATION OF STATIC QUBITS IN THE REGISTERS

In this section, we present the standard characterizations of the static qubits in the registers. They are performed with only two qubits in each registers (either in D1-D2 or D2-D3 for the left (red) register, either in D4-D5 or D5-D6 for the right (blue) register) and slightly different gate-voltage configurations than in the main text. Yet, the metrics presented below in Supplementary Table 1 and Supplementary Table 2 provide valuable insights on the overall qubits' performance.

### A. Larmor and Rabi frequencies

The first two columns of Supplementary Table 1 present the measured Larmor ($f_{\mathrm{L}}$) and Rabi frequencies ($f_{\mathrm{Rabi}}$). On the one hand, we observe that the Larmor frequencies measured for qubits in the left register (red) are consistently higher than those measured for qubits in the right register (blue). This could suggest that the potential and strain fluctuations favor a better alignment between the in-plane principal axes of the $g$-tensors and the external magnetic field for the qubits in the right registers [7]. On the other hand, the Rabi frequencies of qubits in D1 and D6 are lower than those measured in the other quantum dots when normalized to the same driving amplitude (computed as $f_{\mathrm{Rabi}}/A_{\mathrm{D}}$). Qubits in these quantum dots are significantly harder to drive coherently with EDSR. This could arise from the different geometry used for the barrier gates of these quantum dots (see Fig. 1.a from the main text). It could induce a difference in the shape of the confining potential or in the shape of the wavefunction that could cause the driving mechanisms to be less efficient.

### B. Dephasing times

To evaluate the qubits' coherence times, we perform Ramsey and Hahn-echo experiments as respectively sketched in Supplementary Figure 1.a-b. Before the last $X_{\pi/2}$ pulse, we apply a phase shift $\phi = \delta f \cdot t$ (or equivalently a $Z_\phi$ gate) proportional to the free-evolution time $t$ (with $\delta f \approx 0.1 - 2$ MHz fixed) which acts as a virtual detuning. The resulting oscillations are fitted by:

$$A \cos(2\pi \delta f t + \phi_0) \exp(-(t/T_2^{*,\mathrm{H}})^{a_{*,\mathrm{H}}}) + A_0, \tag{2}$$

to extract the corresponding dephasing times $T_2^{*,\mathrm{H}}$ and decay exponents $a_{*,\mathrm{H}}$ which provide insights on the noise experienced by the qubits.

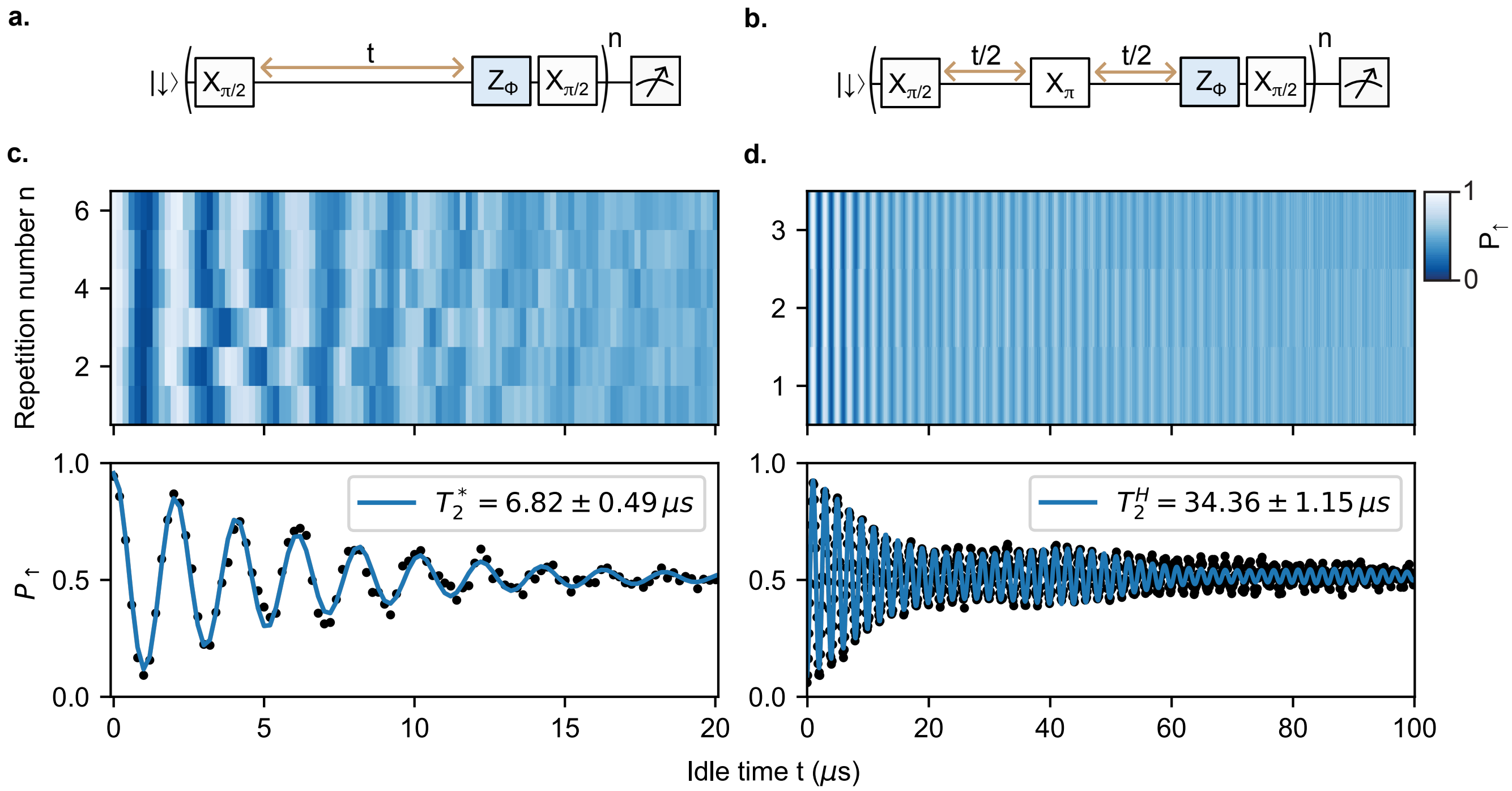

Supplementary Figure 1. **Ramsey and Hahn-echo experiments for the qubit in D6. a.** and **b.** Schematics of the experimental sequence used respectively for the Ramsey and Hahn-echo dephasing-time measurements. The sequence are repeated multiple times successively to improve the reliability of the dephasing-time evaluations. **c.** and **d.** Experimental implementations of the sequences used to evaluate the dephasing times of the qubit in D6. The top panels shows the spin-up probability $P_\uparrow$ as function a function of the idle time $t$ and the repetition number $n$. The bottom panels show the probability averaged over the different repetitions (black dots) and the corresponding fits (blue solid lines) using equations (2) and (3).

To improve the fit reliability, we repeat successively each experiment $n$ times and average the spin-up probability $P_\uparrow$ obtained for the different repetitions as displayed in Supplementary Figure 1.c-d. Measurements are repeated 6 times for the Ramsey and Hahn-echo experiments of qubits located in D3, D4, D5 whilst Hahn-echo experiment in dots D1, D2 and D6 are performed 3 times. Each repetition of the Ramsey sequence takes about a minute thus resulting in a total integration time of about 6 minutes.

In the longest Hahn-echo experiments, marked by a • in Supplementary Table 1, an additional collapse-and-revival is observed and attributed to the hyperfine coupling between the qubit and the spin-9/2 of the $^{73}$Ge nuclei [7–9]. The corresponding oscillations in spin probabilities are thus rather fitted by:

$$\frac{A\cos(2\pi\,\delta f\,t)\exp(-(t/T_2^{\mathrm{H}})^{a_{\mathrm{H}}})}{\left[1-a_0\cos(\pi t B\gamma_{\mathrm{Ge73}})\right]^2}+A_0, \tag{3}$$

with $\gamma_{\mathrm{Ge73}} = 1.49$ MHz/T the gyromagnetic ratio of the germanium spin-9/2 nuclei. The magnetic field $B$ is let as a free parameter.

The extracted dephasing times and decay coefficients are gathered in Supplementary Table 1. We observe that the qubits in D3 and D4, spatially the closest to the conveyor lane, have significantly lower dephasing times. This may be explained by the large tunnel couplings with the quantum dots formed at the extremities of the conveyor lane. Interestingly, we also notice that the qubits in the red register, with the highest Larmor frequencies, have comparable or higher dephasing times than their counterpart in the blue register. Lower Larmor frequencies could suggest an improved alignment of the in-plane $g$-tensor axes, thus a reduced coupling to spinful germanium nuclei and should lead to longer coherence times [7, 10]. The origin of this discrepancy remains unknown and may be related to differences in the qubits' sensitivity to charge noise. Further investigations into the existence of sweet spots as a function of magnetic field orientation [7, 11, 12] could provide additional insights.

| Quantum dot | Larmor frequency $f_L$ (MHz) | Rabi frequency $f_{\text{Rabi}}$ (MHz) | Driving amplitude $A_D$ (mV) | Free-induction dephasing time $T_2^*$ (μs) | Decay exponent $a_*$ | Hahn-echo dephasing time $T_2^H$ (μs) | Decay exponent $a_H$ |
|---|---|---|---|---|---|---|---|
| **D1** | $121.33 \pm 0.01$ | $0.286 \pm 0.001$ | 20 | $7.45 \pm 0.43$ | $1.27 \pm 0.14$ | • $54.27 \pm 1.10$ | $2.08 \pm 0.13$ |
| **D2** | $150.40 \pm 0.01$ | $1.768 \pm 0.002$ | 20 | $9.18 \pm 0.58$ | $1.42 \pm 0.18$ | $36.35 \pm 0.35$ | $1.65 \pm 0.03$ |
| **D3** | $109.01 \pm 0.01$ | $2.041 \pm 0.003$ | 40 | $2.42 \pm 0.04$ | $1.77 \pm 0.07$ | $5.81 \pm 0.05$ | $3.39 \pm 0.13$ |
| **D4** | $23.06 \pm 0.01$ | $0.696 \pm 0.002$ | 25 | $2.74 \pm 0.12$ | $1.82 \pm 0.19$ | $6.40 \pm 0.09$ | $2.97 \pm 0.18$ |
| **D5** | $59.560 \pm 0.004$ | $0.766 \pm 0.001$ | 20 | $5.60 \pm 0.12$ | $1.73 \pm 0.10$ | $9.91 \pm 0.09$ | $2.60 \pm 0.08$ |
| **D6** | $85.404 \pm 0.002$ | $0.163 \pm 0.001$ | 15 | $6.82 \pm 0.49$ | $1.09 \pm 0.11$ | • $34.36 \pm 1.15$ | $1.00 \pm 0.04$ |

Supplementary Table 1. **Characterization of the static qubits in the registers.** Errors bars correspond to one standard deviation from the best fit parameters. We note that different gate-tuning configurations would also result in slightly different measurement values. The Hahn-echo times marked with a • have been obtained from fits using equation (3).

### C. Single-qubit gate fidelity

We finally evaluate single-qubit gate fidelity using Randomized Benchmarking. The Clifford set is generated using the primitives $\{I, X_{\pi/2}, Y_{\pi/2}\}$, giving on average $N_{\text{gate}}$=3.217 primitive operations per Clifford gate [13]. Each qubit is initialized in its spin-down state and subjected to a sequence of $m$ randomly chosen Clifford gates. A final recovery operation is appended so that the overall sequence ideally corresponds to the identity, returning the spin to the down state. This experiment is repeated $n_{\text{var}} = 30$ times. The probability of recovering the down state, $P_\downarrow$, averaged over the $n_{\text{var}}$ repetitions, decays with increasing sequence length as $P_\downarrow = Ap^m + B$, with $p$ the depolarization parameter and $A$, $B$ accounting for initialization and readout errors. From the fitted depolarization parameters, the Clifford gate fidelity $F_C = (1 + p)/2$ and the average native-gate fidelity $F_{1qb} = 1 - (1 - F_C)/N_{\text{gate}}$ can be evaluated. The fidelities $F_{1qb}$ extracted from such measurements are reported in Supplementary Table 2. We did not measured the fidelities for qubits hosted in D1 and D6 as they are much more difficult to drive coherently and as they are only used as readout and initialization ancilla in our experiments.

| Quantum dot | Driving pulse amplitude $A_D$ (mV) | Single-qubit gate fidelity $F_{1qb}$ (%) |
|---|---|---|
| **D2** | 20 | $99.98 \pm 0.003$ |
| **D3** | 40 | $99.98 \pm 0.002$ |
| **D4** | 40 | $99.88 \pm 0.018$ |
| **D5** | 30 | $99.93 \pm 0.011$ |

Supplementary Table 2. **Fidelities of single-qubit gates for static qubits in the registers.** Errors bars correspond to one standard deviation from the best fit parameters.

## Supplementary Section 3. INFLUENCE OF THE CONVEYOR-PULSE PARAMETERS ON COHERENT SPIN SHUTTLING

Here we investigate the effects of the conveyor-pulse amplitudes $A_i$ and frequency $f_c$ on the shuttling process.

### A. Amplitudes of the pulses $A_i$

The choice of conveyor-pulse amplitudes plays a crucial role in enabling coherent shuttling between registers: qubits can only be transported along the shuttling lane if the pulse amplitudes are sufficiently high. High pulse amplitudes are indeed necessary to overcome the local disorder landscape and ensures that the charge remains confined in the minimum of the traveling-wave potential during the shuttling [14, 15].

To investigate this, we perform back-and-forth diabatic spin shuttling experiments with three different pulse amplitudes, as sketched in Supplementary Figure 2.a. We use the shuttling-induced oscillations caused by variations of the local spin quantization axis as a probe to confirm the successful transport of qubits inside the conveyor lane. Once initialized in a $|\downarrow\rangle$ state in D4, the spin is shuttled for a duration $\tau_1$, allowed to freely precess for a time $t$, and then is returned to its initial location for readout. Each experiment is performed with a conveyor frequency $f_c$ set to 50 MHz. We hereafter denote $A$ the pulse amplitude such that $A = A_1 = A_3 = A_2/1.2 = A_4/1.2$ (Supplementary Section 1 C 4). The oscillations measured within the conveyor are displayed in Supplementary Figure 2.b-d respectively for aamplitudes $A$ of 50 mV, 100 mV and 160 mV. The bottom panels display the Fourier transforms highlighting the evolution of the Larmor frequency along the conveyor lane.

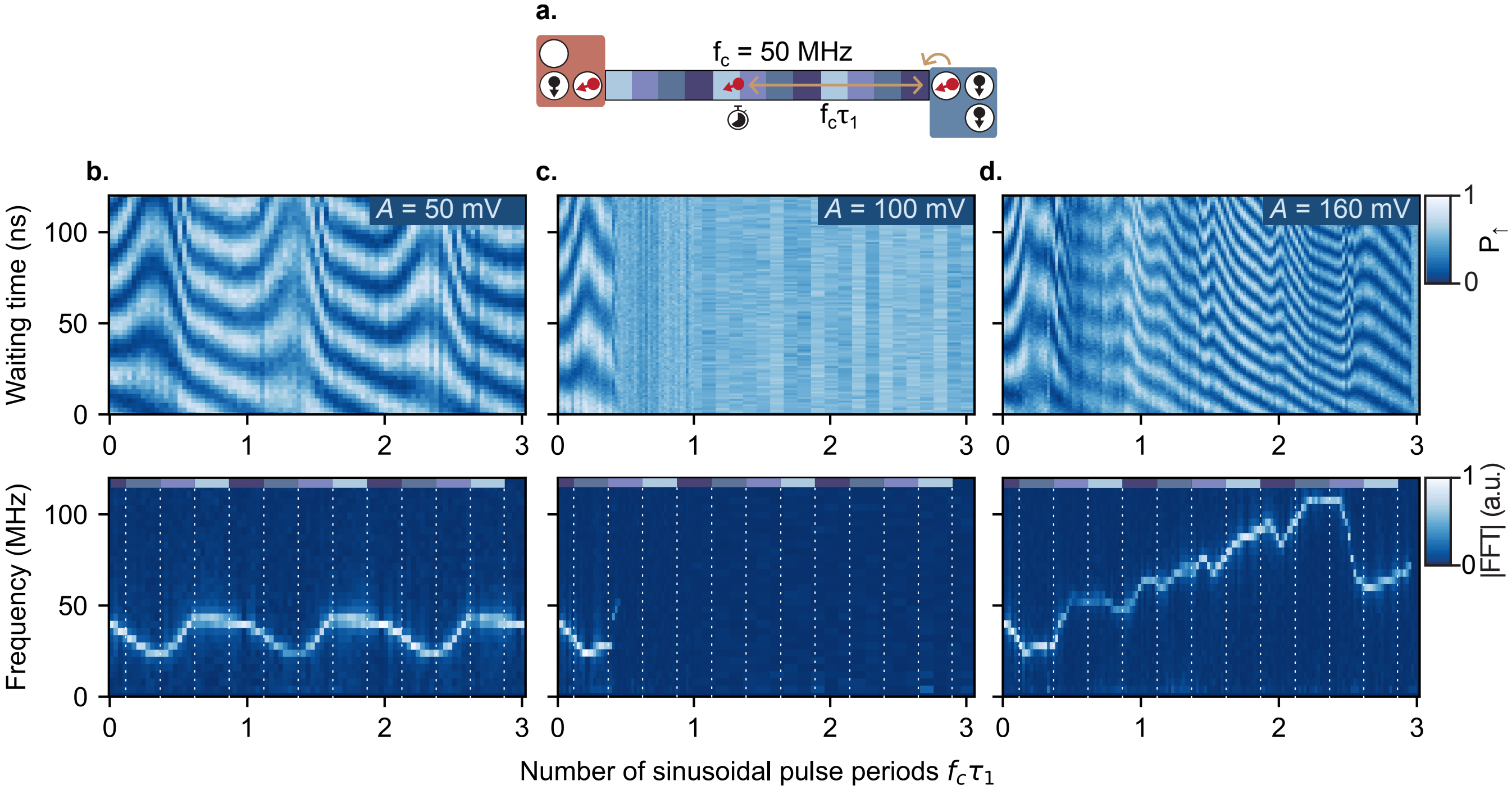

Supplementary Figure 2. **Effect of the conveyor-pulse amplitude on the shuttling. a.** Schematics of the spin transport experiment used to investigate different shuttling regimes depending on the conveyor-pulse amplitude. **b-d.** Shuttling-induced oscillations observed in back-and-forth shuttling as function of the number of periods in the shuttling pulses $f_c\tau_1$. Bottom panels display the corresponding Fourier transforms. The conveyor frequency is fixed $f_c$ at 50 MHz and three different voltage amplitudes are used: 50 mV (b), 100 mV (c), and 160 mV (d).

Supplementary Figure 2.d. illustrates the shuttling-induced rotations measured for an amplitude of $A = 160\,\mathrm{mV}$ equal to the one used for the experiments presented in the main text. This experiment serves as a reference, since we know that, under these conditions, the spin can be reliably transported along the entire conveyor. We observe that the frequency evolves non-monotonically as a function of $f_c\tau_1$ and the FFT is highly similar to the one displayed in Fig 2.k, up to a rotation by 180° and upon rescaling the $y$-axis. The first transformation can be explained by a different shuttling direction, whilst the second one results from shuttling diabatically, i.e. at a higher conveyor frequency ($f_c = 50$ MHz). In this regime, the oscillation frequency is given by the local Larmor frequency rather than by the difference between the local Larmor and the Larmor frequency at the position where the qubit was initialized (D3 for Fig 2.k. and D4 here). Once rescaled, both frequency evolutions for the adiabatic and diabatic regimes are found to be almost identical as shown in Supplementary Figure 6.

For $f_c\tau_1$ close to 3, the oscillations vanish, hinting that the shuttling pulse has become too long. Indeed, for $f_c\tau_1 \simeq 3$, we expect that the potential minimum where the spin was initially loaded is located outside the conveyor lane. Thus, the spin has most likely been ejected into the left register and cannot be returned to D4 through the reversed shuttling pulses.

The Larmor frequency evolution is drastically different when the pulse amplitude is lowered to 50 mV. In this case, it displays a repeating pattern having a characteristic period of $f_c\tau_1 = 1$, as shown in Supplementary Figure 2.b. This means that the qubit experiences a confining potential oscillating with a period corresponding to that of the traveling-wave potential. Considering that these oscillations start from $f_c\tau_1 = 0$ and continue up to $f_c\tau_1 = 3$ included, this indicates that the spin remains localized nearby the right edge of the conveyor channel. In this scenario, the conveyor pulses are insufficient to overcome the local disorder and to displace the qubit along sizable distances. Nevertheless, they still modify the confining potential experienced by the qubit which explains the periodic modulation of the Larmor frequency observed.

Finally, when the amplitude is set to 100 mV (Supplementary Figure 2.c.), an intermediate regime is observed where oscillations are observed up to $f_c\tau_1 \simeq 0.9$ after which they completely disappear. This corresponds to the spin being transported inside the shuttling channel up to certain point and getting lost. Beyond this point, the spin gets trapped or is ejected outside the conveyor due the presence of a local disorder potential maximum.

In a nutshell, the data shown in Supplementary Figure 2 confirm that high pulse amplitudes are required to overcome the local potential disorder and enable shuttling in this device.

### B. Frequency $f_c$

The second key parameter influencing the shuttling dynamics is the conveyor-pulse frequency $f_c$. As mentioned in the main text, this parameter mainly determines the diabaticity of the shuttling with respect to the spin degree of freedom. In this section, we compare the up-state probability evolution across the conveyor lane for two distinct conveyor frequencies (a) $f_c = 50$ MHz and (b) $f_c = 1$ MHz, as displayed in Supplementary Figure 3.

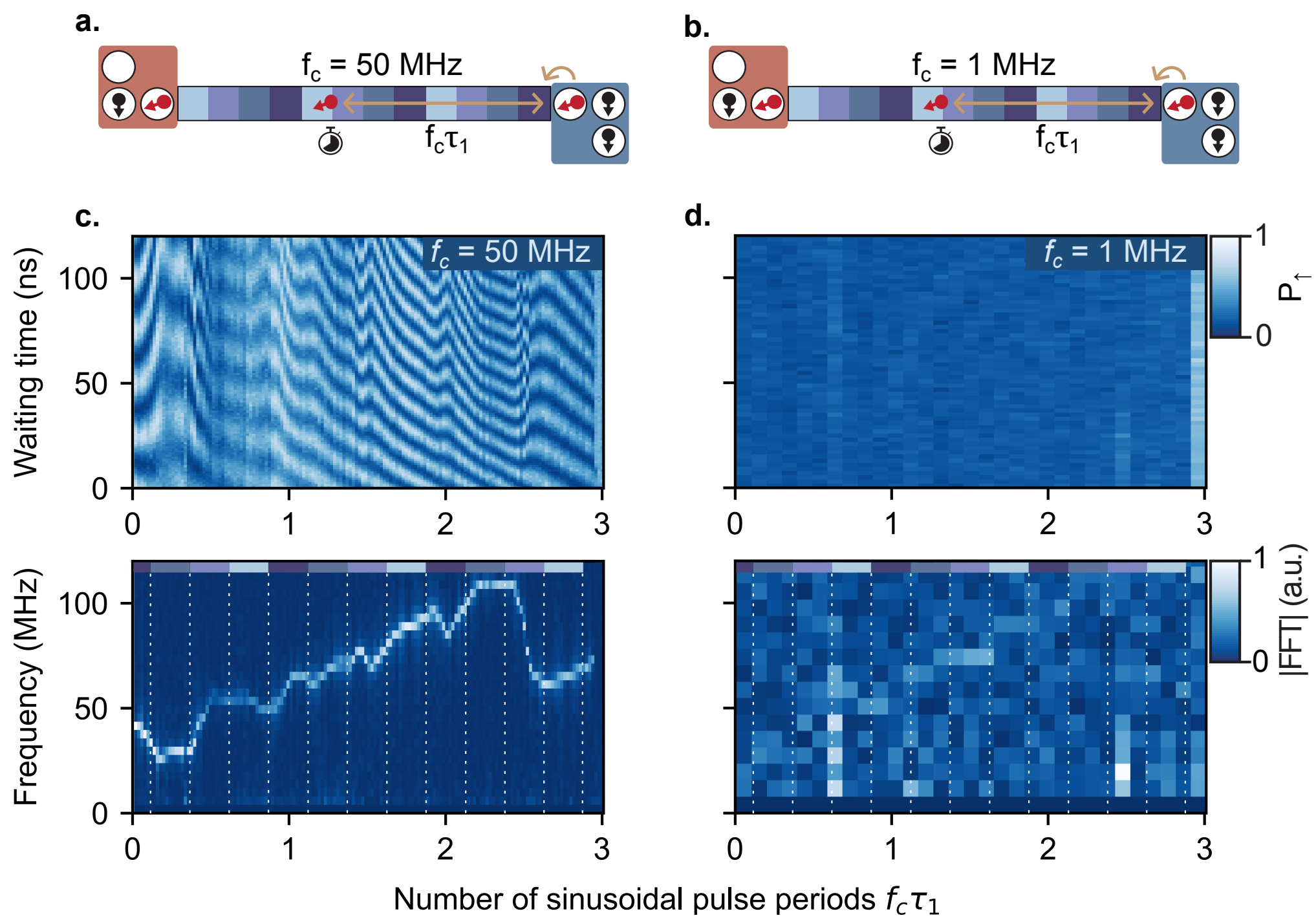

Supplementary Figure 3. **Effect of the conveyor frequency $f_c$ on coherent spin shuttling. a-b.** Schematics of the shuttling experiments performed at different conveyor frequencies $f_c$ of 50 MHz (a) and 1 MHz (b). **c.** Shuttling-induced oscillations measured for a conveyor frequency set to 50 MHz. Bottom panel shows the corresponding Fast Fourier Transform, revealing the qubit Larmor frequency evolution in the shuttling channel. **d.** Absence of shuttling-induced oscillations measured for a conveyor frequency of 1 MHz and the corresponding Fast Fourier Transform. In this regime, the rotations are suppressed, confirming an adiabatic spin transfer during which the spin only accumulates a phase. Both measurements were performed for a conveyor-pulse amplitude of $A = 160$ mV.

Supplementary Figure 3.c manifests the same non-monotonic evolution of the local Larmor frequency as in Supplementary Figure 2.d, understood as arising from rotations induced by the difference in quantization axes. In contrast, Supplementary Figure 3.d exhibits no oscillation pattern when $f_c$ is lowered to 1 MHz. This indicates that the spin transport is in the adiabatic regime where no shuttling-induced spin rotations are triggered and only a phase is accumulated. As a result, a spin initialized in a $|{\downarrow}\rangle$ state will preserve its spin polarization during transport, therefore resulting in the absence of oscillations and an overall low $P_\uparrow$. In this particular case of $f_c = 1$ MHz, the shuttling is actually not completely adiabatic and the spin still experiences residual shuttling-induced rotations. They manifest mainly while driving the qubit inside the conveyor lane (see next section).

As a final observation, we remark that the spin-up probability $P_\uparrow$ suddenly increases when $f_c\tau_1 \simeq 3$. This coincides with the point where the oscillations fade in Supplementary Figure 3.d and thus denotes whenever the conveyor pulses are too long causing the spin to be ejected from the conveyor lane.

## Supplementary Section 4. RESONANT CONTROL AND DEPHASING TIMES WITHIN THE CONVEYOR CHANNEL

In addition to coherent spin transfer, we also demonstrate EDSR control within the shuttling channel using the clavier gates. In this purpose, we perform the experiment display in Supplementary Figure 4.a. A spin initialized and loaded from dot D3 into the conveyor is adiabatically shuttled ($f_{\mathrm{c}} \leq 0.5$ MHz) for a duration $\tau_1$ to an intermediate position. There, a resonant tone, defined by its frequency $f_{\mathrm{D}}$ and duration $t_{\mathrm{D}}$, is applied to one of the clavier gates to coherently drive the qubit. After driving, the qubit is shuttled for a time $\tau_2$ such that it reaches the right qubit register (blue) and is readout by PSB. The clavier gate used to address the qubit is selected to be the most efficient (i.e. to enable the fastest implementation of the $X_{\pi/2}$ gate). Supplementary Figure 4.b to f. display the measured Rabi chevron patterns across the conveyor lane for five different values of $f_{\mathrm{c}}\tau_1$, therefore demonstrating that resonant control can be achieved inside the conveyor lane. We note the presence of additional oscillations in the background of the chevron patterns attributed to residual shuttling-induced oscillations.

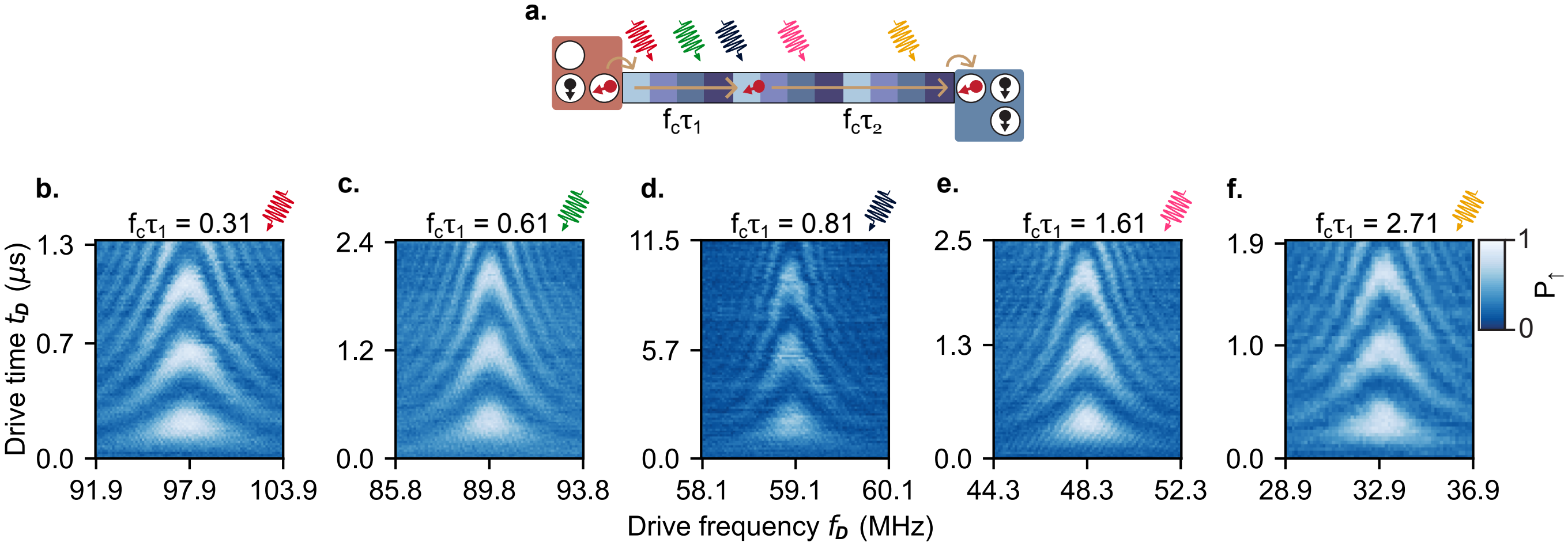

Supplementary Figure 4. **Resonant driving within the shuttling lane. a.** Schematic of the experiments. The shuttling frequency $f_{\mathrm{c}} \leq 0.5$ MHz is chosen to be adiabatic with respect to the spin degree of freedom. **b.-f.** Rabi chevrons measured at different positions within the conveyor lane. The number of sinusoidal pulse periods $f_{\mathrm{c}}\tau_1$ used to reach the position is annotated above each chevron pattern. Colorful arrows in (a) indicate the rough position within the conveyor lane at which these Rabi chevrons have been measured.

We exploit this resonant-driving capability to measure the Larmor frequency $f_{\mathrm{L}}$, as well as the free-induction dephasing time $T_2^*$ and the Hahn-echo dephasing time $T_2^{\mathrm{H}}$ for static qubits inside the conveyor similarly to Supplementary Section 2. For the Ramsey coherence times $T_2^*$, we averaged 6 measurements of about 5 minutes each. For the Hahn-echo decay time $T_2^{\mathrm{H}}$, we averaged between 3 and 6 measurements. All measurement outcomes are gathered in Supplementary Figure 5.

The Larmor frequency $f_{\mathrm{L}}$ displays a non-monotonic evolution with $f_{\mathrm{c}}\tau_1$, nearly identical to the one presented in Supplementary Figure 2.d (up to a reflection along the $y$-axis and a rescaling of the $x$-axis owing to different shuttling direction). All measurements are herein performed at a conveyor frequency of 0.5 MHz, corresponding to a shuttling duration $\tau_{\mathrm{shuttle}} \simeq 5.5$ µs . The free-induction dephasing time $T_2^*$ varies between 1.7 µs and 9.3 µs with a mean value of 3.8 µs and a median value of 3.5 µs. Remarkably, the mean and median $T_2^*$ are significantly shorter than the total time required to shuttle a spin from left to right side of the conveyor. Consequently, the dephasing time of a mobile spin is necessarily longer than the mean dephasing time of a static qubit in the conveyor, suggesting that the mobile qubit is subject to motional narrowing [5, 16–18].

Besides, the Hahn-echo dephasing times $T_2^{\mathrm{H}}$ range from 2.9 µs to 55.9 µs with a mean value of 11.3 µs and a median value of 6.2 µs. We remark that both $T_2^*$ and $T_2^{\mathrm{H}}$ have similar non-monotonic evolutions with $f_{\mathrm{c}}\tau_1$ and show no clear correlation with the Larmor frequency variations. In general, the $T_2^*$ measured in the center of the conveyor ($1 \leq f_{\mathrm{c}}\tau_1 \leq 2$) are longer than those measured nearby the conveyor edges ($0 \leq f_{\mathrm{c}}\tau_1 \leq 1$ and $2 \leq f_{\mathrm{c}}\tau_1 \leq 2.75$). Alternatively, the range of $f_{\mathrm{c}}\tau_1$ where the $T_2^*$ are the longest coincides with the range where the echoing effect is also the most efficient, with an increase of the coherence time by a factor up to 7, as shown in Supplementary Figure 5.c. Although further investigations would be required to conclude on the dependencies observed here, these experiments show that conveyor-mode shuttling can be used to investigate the evolution of qubit properties over (several) micron-long distances.

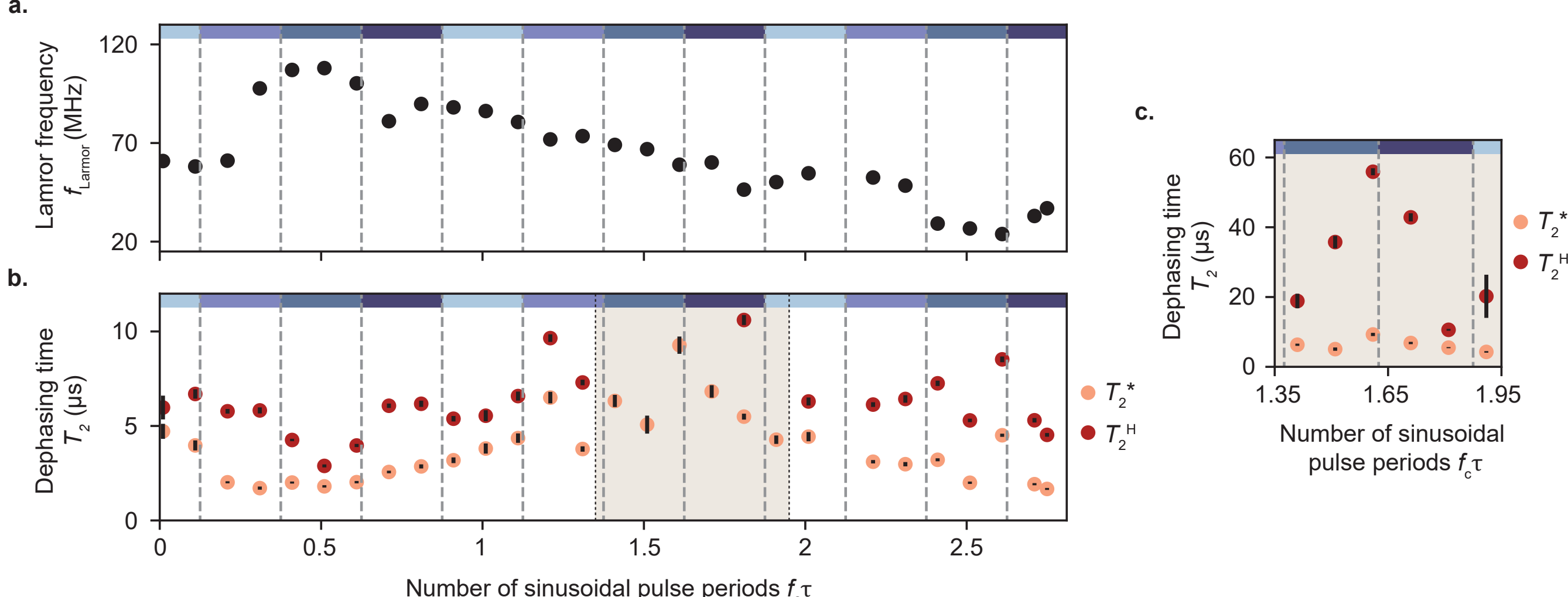

Supplementary Figure 5. **Dephasing times of static qubits measured within the conveyor lane. a.** Larmor frequency evolution $f_L$ at various positions inside the conveyor lane. The qubit position is identified by the number of sinusoidal pulse periods $f_c\tau_1$ used to shuttle. **b.** Free-induction dephasing times $T_2^*$ and Hahn-echo dephasing times $T_2^H$ measured for static qubits inside the conveyor channel. **c.** Zoom-in on the region where the Hahn-echo dephasing times $T_2^H$ is particularly enhanced. Error bars correspond to one standard deviation from the best fit parameters.

## Supplementary Section 5. ROBUSTNESS OF THE LARMOR FREQUENCY EVOLUTION IN THE SHUTTLING CHANNEL

The Larmor frequency evolution along the conveyor channel is highly reproducible. To highlight this, we compare and plot together in Supplementary Figure 6.d, the evolutions measured in experiments presented in Fig. 2.k of the main text, Supplementary Figure 2.d and Supplementary Figure 5.a. For a sake of clarity, we also draw in panels a.-c. schematics of the experiments performed in each case.

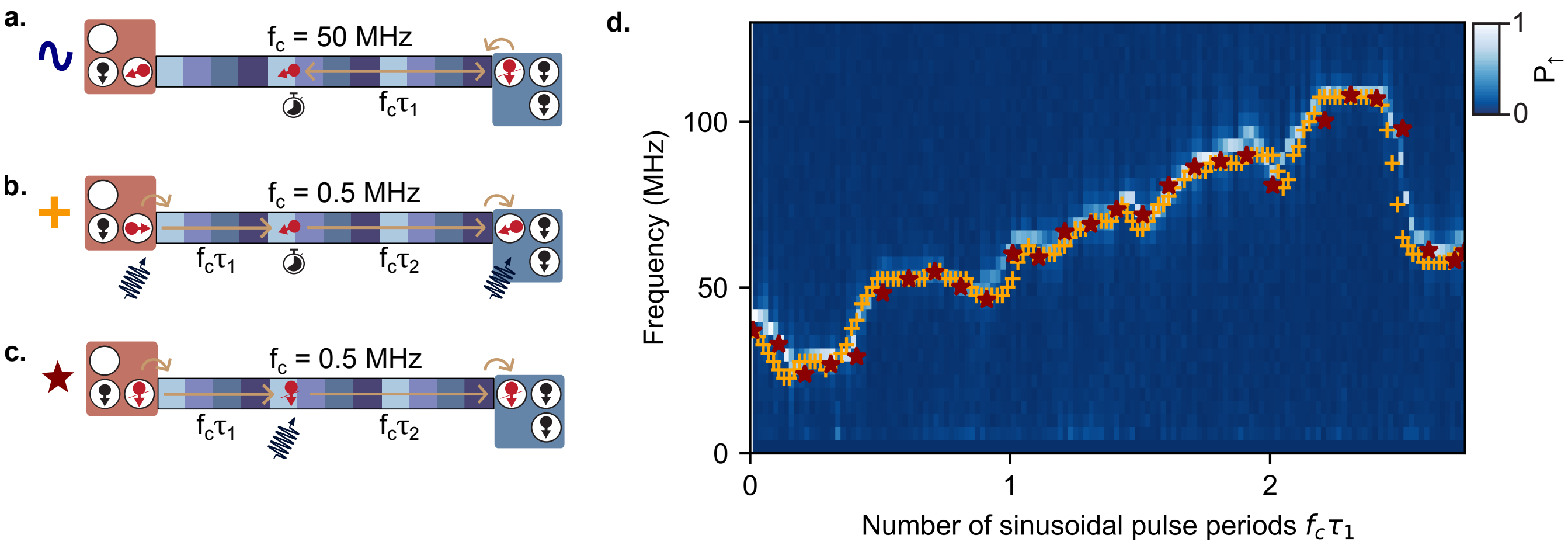

Supplementary Figure 6. **Reproducibility of the Larmor frequency variations between experiments. a.** Schematic of the diabatic shuttling experiment carried out in Supplementary Figure 2.d. **b.** Schematic of the free-precession experiment based on adiabatic shuttling displayed in Fig. 2.k of the main text. **c.** Schematic of the resonant driving experiments shown in Supplementary Figure 5.a. **d.** Comparison of the Larmor frequency variations measured by diabatic transfer (FFT map), free-precession (yellow crosses) and resonant driving (red stars).

In the experiments displayed in Supplementary Figure 2.d, corresponding here to panel (a), the qubit prepared in a $|\downarrow\rangle$ state is diabatically shuttled back-and-forth in the conveyor lane. The coherent oscillations observed in this case arise from changes in the orientation of the quantization axis and their frequency directly corresponds to the local

Larmor frequency.

Alternatively, the experiment of Fig. 2.k (panel (b)), is a free-evolution experiment in which a qubit prepared in a superposition state is adiabatically shuttled ($f_c = 0.5$ MHz) from the left register (D3) to the right one (in D4). The observed oscillation frequency corresponds to the difference between the initial Larmor frequency in D3, $f_L^{D3} \simeq 110$ MHz, and the Larmor frequency at the position reached after conveyor pulses of duration $\tau_1$, denoted $f_L(\tau_1)$. Orange crosses reported in Supplementary Figure 6.d. mark the measured Larmor frequencies across the conveyor lane.

Finally, in Supplementary Figure 5.a (corresponding to the schematic in panel (c)), the Larmor frequencies are directly measured by EDSR, where the qubit is initially prepared in a basis $|\downarrow\rangle$ state, driven within the conveyor lane and shuttled adiabatically from left to right register. Red stars highlight the Larmor frequency evolution reported in the previous section.

It is important to note that the shuttling process in Supplementary Figure 2.d starts from dot D4 whereas, in the other two experiments, the qubit is initially in dot D3. In order to account for the difference in shuttling direction, we apply the transformation $f_c\tau_1 \rightarrow 2.75 - f_c\tau_1$ for the cases where the shuttling is initially hosted in D3.

All Larmor frequency evolutions from Fig. 2.k in the main text, Supplementary Figure 6.d. and Supplementary Figure 5.a, show a remarkable agreement, proving that the Larmor frequency evolution in the conveyor lane is highly reproducible between experiments and does not depend on the shuttling direction.

## Supplementary Section 6. RESONANT OPERATION IN MULTIPLE ROTATING FRAMES

In this work, some qubits are operated in multiple sites associated with different local Larmor frequencies and thus different rotating frames. The aim of this section is to describe how to properly operate qubits resonantly in such conditions.

### A. Rotating and laboratory frames

In experiments where spin qubits are controlled resonantly, the operations are typically described in the rotating frame (RF) that follows the qubit precession in the laboratory frame (LF). Rotations around axes laying on the equatorial plane of the Bloch sphere, such as $X_\theta$ and $Y_\theta$ gates, are implemented by applying resonant pulses. The duration of the pulse determines the angle of the rotation while the axis of rotation is set by the relative phase between the driving field and the instantaneous phase of the qubit precession $\phi_L = 2\pi f_L t$ that serves as a reference.

Consequently, when operating qubits in the rotating frame, the rotations around the $z$-axis of the Bloch sphere or equivalently the $Z$ gates, are typically applied "virtually" by changing the reference phase for the subsequent driving pulses rather than by applying any physical operation to the qubit. Virtual $Z$ gates are thus instantaneous and leave the qubit state unchanged in the laboratory frame but instead redefine the rotating frame as shown in Supplementary Figure 7. This contrasts fundamentally with operations in the laboratory frame based on real $Z$ gates, where we let the qubit idling for a finite time to accumulate a phase through its precession.

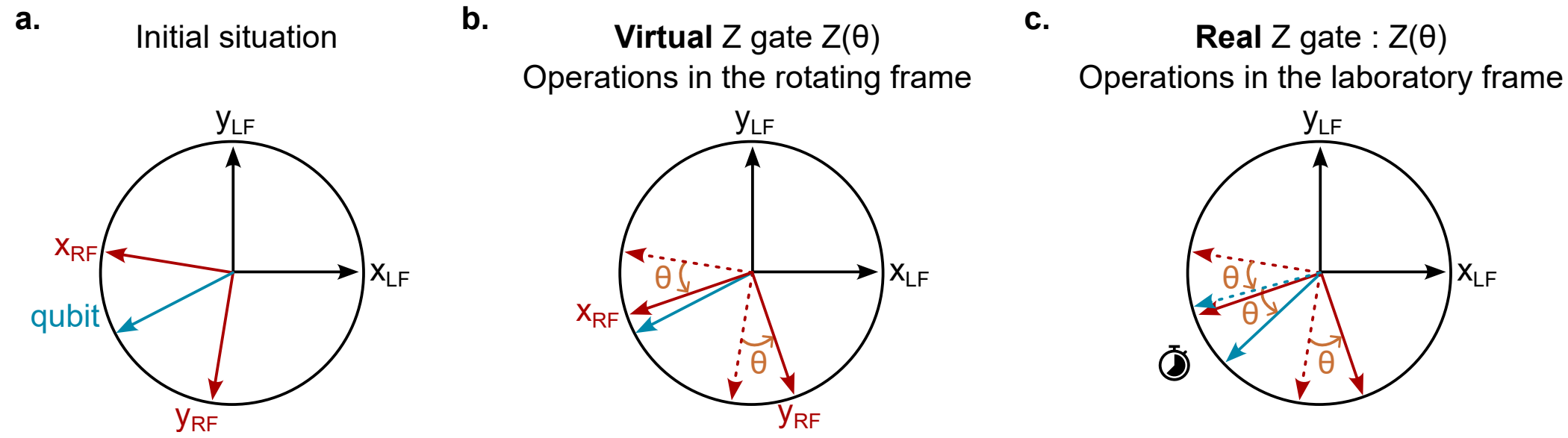

Supplementary Figure 7. **Comparison between real and virtual Z gates.** In these schematics, the direction of the rotations induced by virtual $Z$ gates follows the convention used in our control software.

Using virtual $Z$ gates implies to have a control system that keeps track of the phase of the first resonant pulse applied and calculates the relative phases of all the following pulses. The system must take into account the time elapsed between the pulses but also potential reference-phase updates associated with virtual $Z$ gates. For a static qubit having a well-defined Larmor frequency, the phase tracking can be easily implemented by following the instantaneous phase $\phi_{\text{ref}}(t) = 2\pi t f_L$ of a dummy signal oscillating at $f_L$ throughout the experiment. In the absence of phase updates for the reference signal, a microwave pulse applied at time $t'$ with phase $\phi = \phi_{\mathrm{ref}}(t') + \phi_0$ leads to a rotation along an axis in the $x$-$y$ plane of the Bloch sphere titled by an angle $\phi_0$ with respect to the $x$-axis.

Before moving forward, we note that in our control software ("pulse-lib"), virtual $Z(\theta)$ gates are implemented by rotating the axes of the rotating frame by an angle $\theta$ in the direction of their time evolution as displayed in Supplementary Figure 7. From the qubit's perspective, this is equivalent to a phase shift by $-\theta$. Despite this, we loosely refer to this operation as $Z(\theta)$ gate throughout the manuscript.

### B. Synchronization and tracking of rotating frames for a moving qubit

The above phase tracking protocol needs to be adapted when working with a mobile qubit operated at different locations associated with distinct local Larmor frequencies, hence distinct rotating frames.

To illustrate this, we focus on the simplified case of a spin qubit shuttled between two sites A and B, defined by their respective Larmor frequency $f_{\mathrm{L}}^{\mathrm{A}}$ and $f_{\mathrm{L}}^{\mathrm{B}}$. We hereafter denote the axes of the rotating frames of site A (resp. B) by $x_{\mathrm{RF}}^{\mathrm{A}}$ and $y_{\mathrm{RF}}^{\mathrm{A}}$ (resp. $x_{\mathrm{RF}}^{\mathrm{B}}$ and $y_{\mathrm{RF}}^{\mathrm{B}}$) and the axes of the laboratory frame by $x_{\mathrm{LF}}$ and $y_{\mathrm{LF}}$. We consider an experiment where we aim at applying two successive $X_{\pi/2}$ gates to the qubit, one in site A and one in site B, interleaved by a shuttling step. To simplify the discussion, we assume that the transfer from site A to site B is fully adiabatic (thus it does not lead to any net qubit rotation) and that the phase accumulated during the transfer is a multiple of $2\pi$. A more general and realistic case is discussed in the next section.

When the experiment starts at $t = 0$, the rotating frames of A and B are both synchronized with the laboratory frame (see step (0) in the bottom panel of Supplementary Figure 8). At this point, a Larmor frequency is assigned to each site. This defines the evolution speeds of the local rotating frames and sets the reference frequencies used for the phase tracking in each site.

In site A, the qubit is first subjected to a resonant pulse generating a rotation along the $x_{\mathrm{RF}}^{\mathrm{A}}$ axis. Once shuttled into site B at time $t = t_1$, the qubit is subjected to another resonant pulse (step (1)). If nothing is done, the local phase tracking will cause the pulse to be applied with an initial phase $2\pi t_1 f_{\mathrm{L}}^{\mathrm{B}}$ such that it corresponds to a rotation around the $x_{\mathrm{RF}}^{\mathrm{B}}$ axis. Since $f_{\mathrm{L}}^{\mathrm{A}} \neq f_{\mathrm{L}}^{\mathrm{B}}$, the $x_{\mathrm{RF}}^{\mathrm{A}}$ and $x_{\mathrm{RF}}^{\mathrm{B}}$ axes will generally not coincide at time $t_1$. As a result, the application of the two resonant pulses will not be equivalent to the application of two successive $X_{\pi/2}$ gates on the qubit.

To tackle this issue, both rotating frames must be aligned before applying the second pulse in site B. To this aim, a virtual $Z$ gate by an angle $\theta_{\mathrm{sync}} = 2\pi t_1(f_{\mathrm{L}}^{\mathrm{A}} - f_{\mathrm{L}}^{\mathrm{B}})$ can be applied in the rotating frame B just before the second pulse (see Supplementary Figure 8 step (2)). This correction will ensure that the second pulse is applied along the same axis as the first one. After this correction, all subsequent pulses applied in site B will perform the intended operations thanks to the standard phase tracking at $f_{\mathrm{L}}^{\mathrm{B}}$.

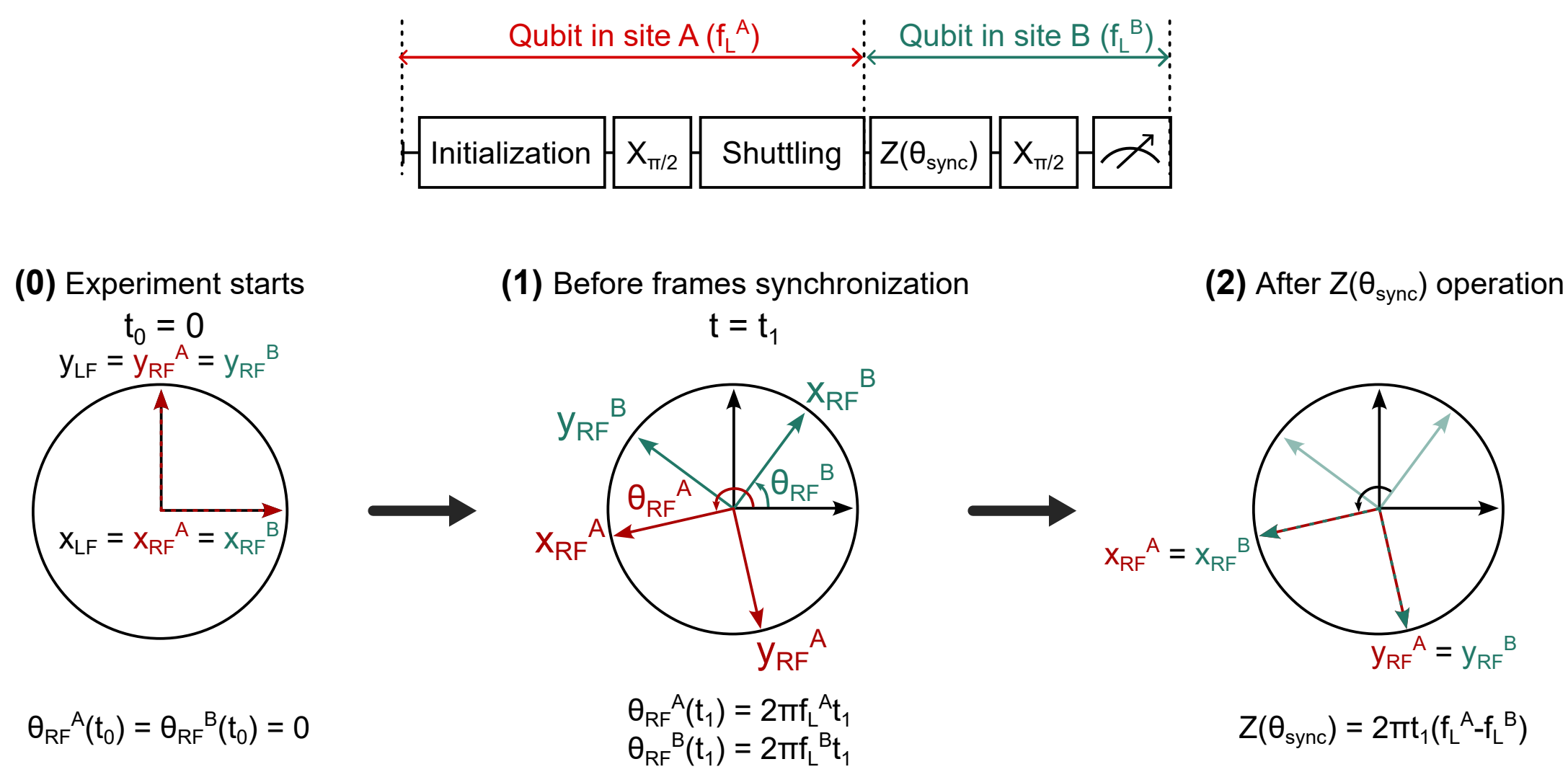

Supplementary Figure 8. **Synchronization of rotating frames across two sites.** Top panel: Schematic of the quantum circuit considered where we intend to perform two $X_{\pi/2}$ gates in different site A and B having their own local Larmor frequency $f_{\mathrm{L}}^{\mathrm{A}}$ and $f_{\mathrm{L}}^{\mathrm{B}}$. Bottom panel: Evolution of the rotating frames during the experiment. At time $t = t_1$, the frames are not synchronized as $f_{\mathrm{L}}^{\mathrm{A}} \neq f_{\mathrm{L}}^{\mathrm{B}}$. A virtual Z rotation by an angle $\theta_{\mathrm{sync}} = 2\pi t_1(f_{\mathrm{L}}^{\mathrm{A}} - f_{\mathrm{L}}^{\mathrm{B}})$ allows to resynchronize the two rotating frames and thus to apply the second pulse along the same axis than the first one.

Although this example provides insights on how to combine multiple rotating frames for a shuttled qubit, we would like to emphasize two additional points that are particularly relevant for the entanglement-distribution experiments presented Fig. 4 of the main text. First, a similar phase correction is also necessary in the case of static qubits having a Larmor frequency that changes during the experiment. This applies to the static qubit in D2, whose frequency is modified after shuttling of the mobile qubit due to the modification of the electrostatic environment. Second, when a qubit is shuttled between sites A and B, any phase update done in site A should be duplicated into site B. This is relevant for the single-qubit phase corrections that are applied following the exchange pulse in the implementation of a two-qubit $CZ$ gate.

## Supplementary Section 7. COMBINING SHUTTLING-INDUCED ROTATIONS WITH RESONANT CONTROL

The purpose of this section is to describe the protocols necessary to combine diabatic spin shuttling and resonant control, as done in Fig. 3 and Fig. 4 of the main text.

During a diabatic spin shuttle, the strong spin-orbit coupling results in a rotation of the spin state, denoted as $R_{\text{shut}}$. This rotation is more naturally described in the laboratory frame where its rotation axis and angle are fixed. The latter depends on multiple parameters including the shuttling duration and speed, the details of the local variations of the quantization axes along the shuttling path but also the idling times between the different pulses of the sequence.

Once the shuttling is performed, $R_{\text{shut}}$ can be compensated by applying a recovery-gate operation $R_{\text{rec}}$ that compensates for the shuttling-induced rotation, such that $R_{\text{rec}} = R_{\text{shut}}^{-1}$. The recovery gate is applied using a resonant pulse and thus is more naturally described in the qubit's rotating frame. In this frame, it can be decomposed in three successive rotations such that $R_{\text{rec}} = Z_\gamma X_\beta Z_\alpha$ or equivalently as:

$$\begin{aligned} R_{\text{rec}} &= Z_\gamma X_\beta Z_\alpha \\ &= Z_{\alpha+\gamma} Z_{-\alpha} X_\beta Z_\alpha \\ &= Z_{\alpha+\gamma} X_\beta^{-\alpha}. \end{aligned} \tag{4}$$

The later decomposition shows that the recovery operation can be decomposed in two main operations. The first term $X_\beta^{-\alpha}$ can be interpreted as a rotation by an angle $\beta$ around an axis titled by $-\alpha$ with respect to the $x_{\text{RF}}$-axis of the rotating frame. The second term $Z_{\alpha+\gamma}$ is a pure phase correction.

As $R_{\text{shut}}$ and $R_{\text{rec}}$ are more conveniently described in different reference frames and thus are more naturally compatible with different types of operations, specific measures have to be taken to combine both. In the following, we detail these measures for the case of one or multiple cycles including diabatic shuttling and recovery operation.

### A. Single shuttle case

We first focus on the single shuttle case, relevant for the experiments shown in Fig 3.d and Fig. 4 of the main text. We consider a spin initially localized at site A that is diabatically shuttled to site B, undergoing shuttling-induced rotations when transferred. As mentioned above, $R_{\text{shut}}$ corresponds to a rotation around a fixed axis $\hat{n}$ in the laboratory frame. In the rotating frame, $\hat{n}$ rotates around the $z$-axis at the Larmor frequency. Thus, in this frame, the exact rotation performed depends on the time $t_{\text{init}}$ elapsed between the beginning of the experiment and the beginning of the shuttling operation. If this time is changed, both $R_{\text{shut}}$ and $R_{\text{rec}}$ will be different.

Consequently, if we want to be able to define a unique recovery operation that is independent of the initialization procedure and of the exact sequence of operations performed on the qubits before shuttling, $t_{\text{init}}$ must remain constant. We achieved this by adding a large variable buffer time at the very beginning of the sequence that we adapt to ensure that $t_{\text{init}}$ is fixed across all experiments.

We note that in our work, this single shuttle case is relevant for two types of experiments: (1) experiments where the spin is being shuttled back-and-forth once in the conveyor lane and brought back to its initial quantum dot and (2) experiments where the spin is shuttled in only one direction, such that it ends up being in another quantum dot. In the first scenario, as the spin returns to its initial location, the rotating frames before and after shuttling are the same. Thus, no specific resynchronization protocol is required and the shuttling-induced rotation can be directly compensated by the recovery gate. In the second case where the spin ends in a different site, a synchronization of the two rotating frames (from site A to B) is needed. As previously mentioned in Supplementary Section 6, this frame realignment boils down to adding a virtual $Z(\theta_{\text{sync}})$ gate. The total recovery operation after shuttling is then $Z_\gamma X_\beta Z_\alpha Z(\theta_{\text{sync}})$ and thus has exactly the same form up a redefinition of the recovery-gate angles.

### B. Multiple successive shuttling cycles

This section focuses on the case presented in Fig. 3 in the main text, in which the spin, initially in site A, is diabatically shuttled back-and-forth $N$ times and returned after each cycle to its initial position. After each round trip, lasting for a duration $t_{\mathrm{shuttle}}$, the spin is subjected to the same recovery gate compensating for shuttling-induced rotations but also to an additional real $Z$ gate. In a nutshell, this extra real $Z$ gate, consisting in an idling stage of duration $t_{\mathrm{wait}}$, ensures that the spin state in the laboratory frame is always the same at the beginning of each shuttling cycle.

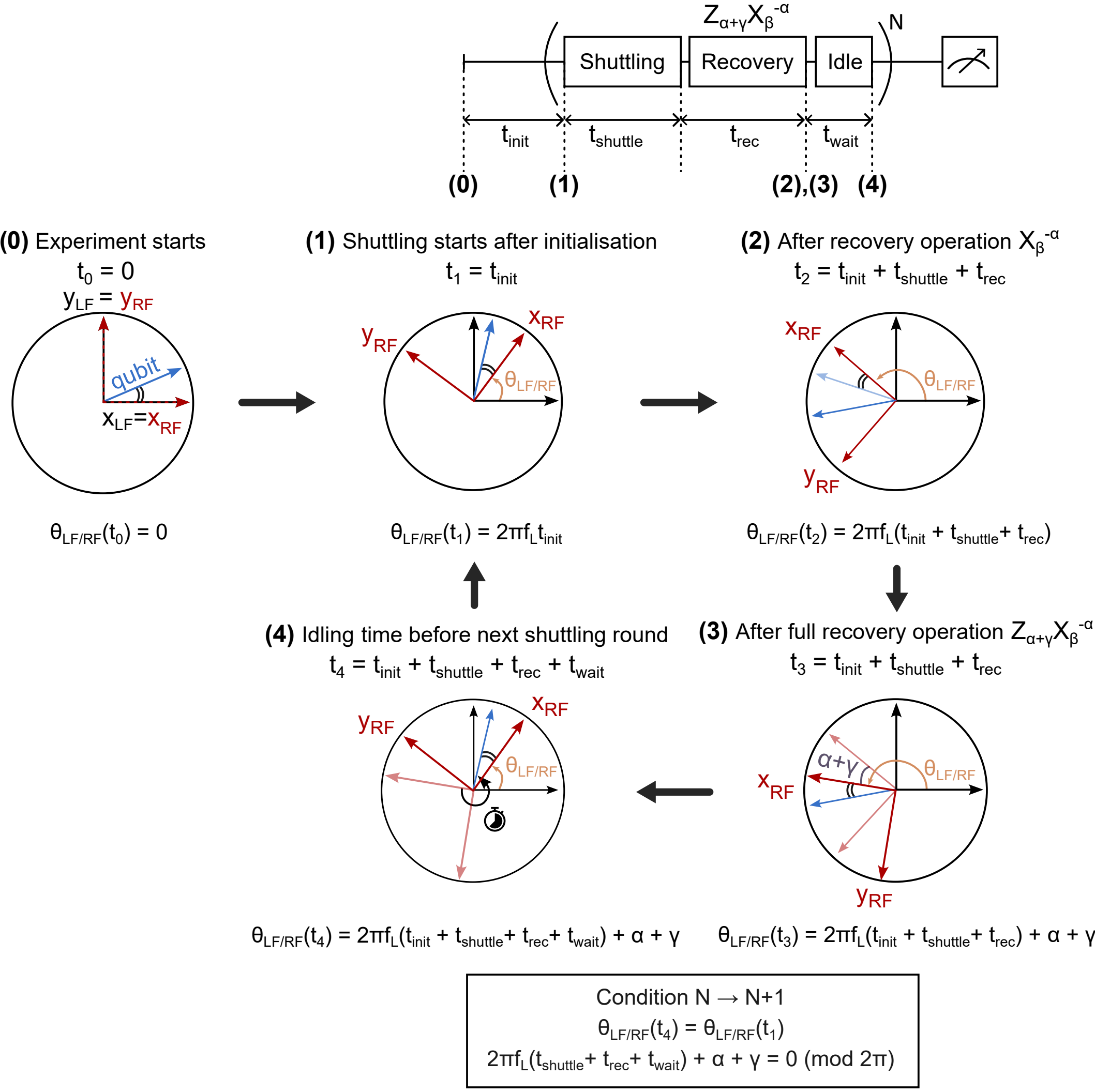

Supplementary Figure 9. **Frame synchronization for multiple shuttling cycles.** Top panel: schematic of the quantum circuit performed when shuttling diabatically $N$ times. Bottom panel: Evolution of the qubit's state vector in both the rotating frame and laboratory frame. For clarity, the state vector is represented as projected onto the equator of the Bloch sphere.

Supplementary Figure 9 **(0) - (4)** depicts the evolution of the qubit (blue) and its rotating frame (red) with respect to the laboratory frame (black) for all steps involving $N$ shuttling cycles. For a sake of clarity and without loss of generality, the schematic is reduced to the equatorial plane of the Bloch sphere and we assume that the qubit has initially a finite phase with respect to the laboratory frame. Top panel is a simplified version of the quantum circuit performed in Fig. 3 of the main text where we define the relevant timescales.

**(0):** At $t_0 = 0$ the experiment begins and all the axes of the laboratory frame are aligned with those of the rotating frame, hence $x_{\mathrm{LF}} = x_{\mathrm{RF}}$ and $y_{\mathrm{LF}} = y_{\mathrm{RF}}$.

**(1):** At $t_1 = t_{\mathrm{init}}$, the qubit is loaded into the conveyor lane. All along the initialization and state preparation procedure, the qubit precessed at its Larmor frequency. Thus at time $t_{\mathrm{init}}$, the rotating frame is detuned with respect to the laboratory frame by an angle $\theta_{\mathrm{LF/RF}}(t_1) = 2\pi f_{\mathrm{L}} t_{\mathrm{init}}$.

**(2):** The spin is diabatically shuttled back-and-forth during a time $t_{\mathrm{shuttle}}$ and meanwhile the rotating frame precesses. Once returned, the spin undergoes a $X_{\beta}^{-\alpha}$ rotation due to the recovery pulse (equation (4)) that takes a duration $t_{\mathrm{rec}}$. After this pulse, the rotating frame is detuned by $\theta_{\mathrm{LF/RF}}(t_2) = 2\pi f_{\mathrm{L}}(t_{\mathrm{init}} + t_{\mathrm{shuttle}} + t_{\mathrm{rec}})$ with respect to the lab frame.

**(3):** The second part of the recovery gate $Z_{\alpha+\gamma}$ is applied (phase update for all subsequent pulses), resulting in changes of the rotating frame axes by an angle $\alpha + \gamma$ with respect to the laboratory frame. At this stage, in the rotating frame, the rotations induced by the fast shuttling have been completely undone. However, in the laboratory frame, both the qubit state and the axes of the rotating frame are not yet identical from those at step **(1)**.

**(4):** In a final step, we let the qubit freely precesses for a duration $t_{\mathrm{wait}}$ in the quantum dot until its state in the laboratory frame is identical to its state at step **(1)** and until the rotating frame axes come back to the positions they had at step **(1)**. This idling stage ensures that the $N^{\mathrm{th}}$ shuttling round will affect the spin the same way as the $(N-1)^{\mathrm{th}}$ cycle. Whenever the situation in **(4)** is equivalent to the situation in **(1)**, the qubit can be shuttled once again so that $N \rightarrow N+1$.

From this description, we can finally derive the following resynchronization condition:

$$\theta_{\mathrm{LF/RF}}(t_4) = \theta_{\mathrm{LF/RF}}(t_1) \pmod{2\pi}, \tag{5}$$

$$2\pi f_{\mathrm{L}}(t_{\mathrm{init}} + t_{\mathrm{shuttle}} + t_{\mathrm{rec}} + t_{\mathrm{wait}}) + \alpha + \gamma = 2\pi f_{\mathrm{L}}\, t_{\mathrm{init}} \pmod{2\pi}, \tag{6}$$

$$2\pi f_{\mathrm{L}}(t_{\mathrm{shuttle}} + t_{\mathrm{rec}} + t_{\mathrm{wait}}) + \alpha + \gamma = 0 \pmod{2\pi}. \tag{7}$$

From the above equation 7, we can evaluate the idling duration to be:

$$t_{\mathrm{wait}} = -\frac{\alpha+\gamma}{2\pi f_{\mathrm{L}}} - t_{\mathrm{rec}} - t_{\mathrm{shuttle}} \pmod{1/f_{\mathrm{L}}}. \tag{8}$$

Equation 8 reveals that the additional idling time stems from two contributions: the first term results from the phase accumulated during the shuttling process, whilst the second term accounts for the precession of the rotating frame with respect to the laboratory frame occurring during the duration of a cycle.

## Supplementary Section 8. QUANTUM STATE TOMOGRAPHY (QST) OF A TWO-QUBIT SYSTEM AFTER DIABATIC SPIN SHUTTLING

### A. Introduction to QST

To evaluate the quality of the quantum information transfer and the entanglement distribution, we perform quantum state tomography (QST). The latter allows to reconstruct the density matrix $\rho$ of multi-qubit system by combining multiple measurement outcomes.

In general, the density matrix $\rho$ of a two-qubit system $\mathrm{Q_A}$, $\mathrm{Q_B}$ can be written as $\rho = \sum_{i=1}^{16} c_i M_i$, with $M_i$ being a set of linearly-independent measurement operators which span the density-matrix space. For example, the $M_i$ can be the two-qubit Pauli operators given by $\sigma_{\mathrm{A}} \otimes \sigma_{\mathrm{B}}$ where $\sigma_{\mathrm{A,B}} \in \{\mathbb{1}, \sigma_x, \sigma_y, \sigma_z\}$ are the single-qubit Pauli matrices and $\otimes$ denotes the tensor product between qubit A and qubit B subspaces. QST consists in computing the coefficients $c_i$ for a given set of $M_i$ based on their expectations values $m_i = \Sigma_{j=1}^{16} \mathrm{Tr}(M_i M_j) c_j$. The latter can be determined by measuring the joint probabilities in various multi-qubit readout bases. Instead of directly solving the linear system relating the expectation values $m_i$ to the $c_i$, the coefficients $c_i$ are typically evaluated by maximum likelihood estimation. This ensures that the reconstructed $\rho$ satisfies all physical requirements for a density matrix, namely it is a Hermitian positive semi-definite matrix with a unit trace [19–23].

The fidelity of the measured state $\rho$ to an ideal density matrix, denoted $\rho_{\mathrm{target}}$, is given by:

$$F(\rho, \rho_{\mathrm{target}}) = \left(\mathrm{Tr}\sqrt{\sqrt{\rho_{\mathrm{target}}}\,\rho\,\sqrt{\rho_{\mathrm{target}}}}\right)^2. \tag{9}$$

The above definition is symmetric in its argument $\rho_{\mathrm{target}}$ and $\rho$ [24] and it simplifies to $F(\rho, \rho_{\mathrm{target}}) = \mathrm{Tr}(\rho_{\mathrm{target}}\varphi) = \langle\varphi|\,\rho\,|\varphi\rangle$ if $\rho_{\mathrm{target}} = |\varphi\rangle\langle\varphi|$ is a pure state.

### B. About the QST implementation

In this work, the system comprises a static qubit $\mathrm{Q_A}$ located in D2 and the moving qubit $\mathrm{Q_B}$ shuttled diabatically from D3 to D4.

We first prepare the system in the target state in the left (red) register, then shuttle $Q_B$ to the right (blue) register, apply the recovery operation in D4 and measure the two-qubit state in a single-shot fashion. The joint spin-state probabilities, $P_{\downarrow\downarrow}$, $P_{\downarrow\uparrow}$, $P_{\uparrow\downarrow}$ and $P_{\uparrow\uparrow}$ are determined by averaging the measurement outcomes over 5000 repetitions.

To change the measurement basis, we apply before readout either $\mathbb{1}$ (identity), $X_{\pi/2}$ or $Y_{\pi/2}$ single-qubit gates on each qubit similarly to refs. [25–28]. This allows to map the projections of the qubits' states along the $x$-axes or $y$-axes onto projections along the $z$-axes, that are the only directly measurable quantities. In total, we perform 9 successive measurements with single-qubit pre-rotations chosen in the following ensemble:

$$\{\mathbb{1}^{\mathrm{A}}\otimes\mathbb{1}^{\mathrm{B}},\ \mathbb{1}^{\mathrm{A}}\otimes X^{\mathrm{B}}_{\pi/2},\ \mathbb{1}^{\mathrm{A}}\otimes Y^{\mathrm{B}}_{\pi/2}, X^{\mathrm{A}}_{\pi/2}\otimes\mathbb{1}^{\mathrm{B}},\ X^{\mathrm{A}}_{\pi/2}\otimes X^{\mathrm{B}}_{\pi/2},\ X^{\mathrm{A}}_{\pi/2}\otimes Y^{\mathrm{B}}_{\pi/2}, Y^{\mathrm{A}}_{\pi/2}\otimes\mathbb{1}^{\mathrm{B}},\ Y^{\mathrm{A}}_{\pi/2}\otimes X^{\mathrm{B}}_{\pi/2},\ Y^{\mathrm{A}}_{\pi/2}\otimes Y^{\mathrm{B}}_{\pi/2}\}. \quad (10)$$

Overall, this yields to $9\times4=36$ joint spin-state probabilities and to $9\times3=27$ independent parameters that we used to determine the 16 coefficients of the density matrix.

### C. Evaluation of the state preparation and measurement errors (SPAM)

The density matrices reconstructed using QST are sensitive to state preparation and measurement (SPAM) errors which degrade the final fidelity. In the experiments presented in Fig. 4 of the main text, the preparation and readout of both qubits $Q_A$ and $Q_B$ involve three initialization steps (pairs D1–D2, D2–D3 and D5–D6) and two projective measurements (pairs D1–D2 and D4–D5). The errors accumulated across these processes degrade the reconstructed density matrix, leading to a computed fidelity that may be noticeably lower than the true fidelity achievable in the system.

Following refs. [25, 26], we introduce a SPAM matrix, denoted $M_{\mathrm{SPAM}}$, to account for the aforementioned errors and provide a fidelity estimate assuming that these errors are absent. The measured probabilities can now be converted into intrinsic ones (with no SPAM errors) as $P_{\mathrm{meas}}=M_{\mathrm{SPAM}}P_{\mathrm{intr}}$. Knowing $M_{\mathrm{SPAM}}$ and $P_{\mathrm{meas}}$, one can then perform the QST analysis on $P_{\mathrm{intr}}=M^{-1}_{\mathrm{SPAM}}P_{\mathrm{meas}}$ to reconstruct the density matrix and evaluate the intrinsic fidelity. Hereafter, we denote the non-corrected and corrected fidelities by $F$ and $F_{\mathrm{cor}}$.

We evaluate the SPAM matrix by measuring the two-spin state probabilities of $Q_A$ and $Q_B$ after shuttling and recovery operation with the system initially prepared in one of the basis states. An example of such SPAM matrix is shown in Supplementary Figure 10. Because of procedure used, some errors in the recovery gate (meant to undo the shuttling-induced rotations) are also included in the SPAM matrix. These recovery-gate calibration errors cannot be disentangled from the state preparation and measurement errors.

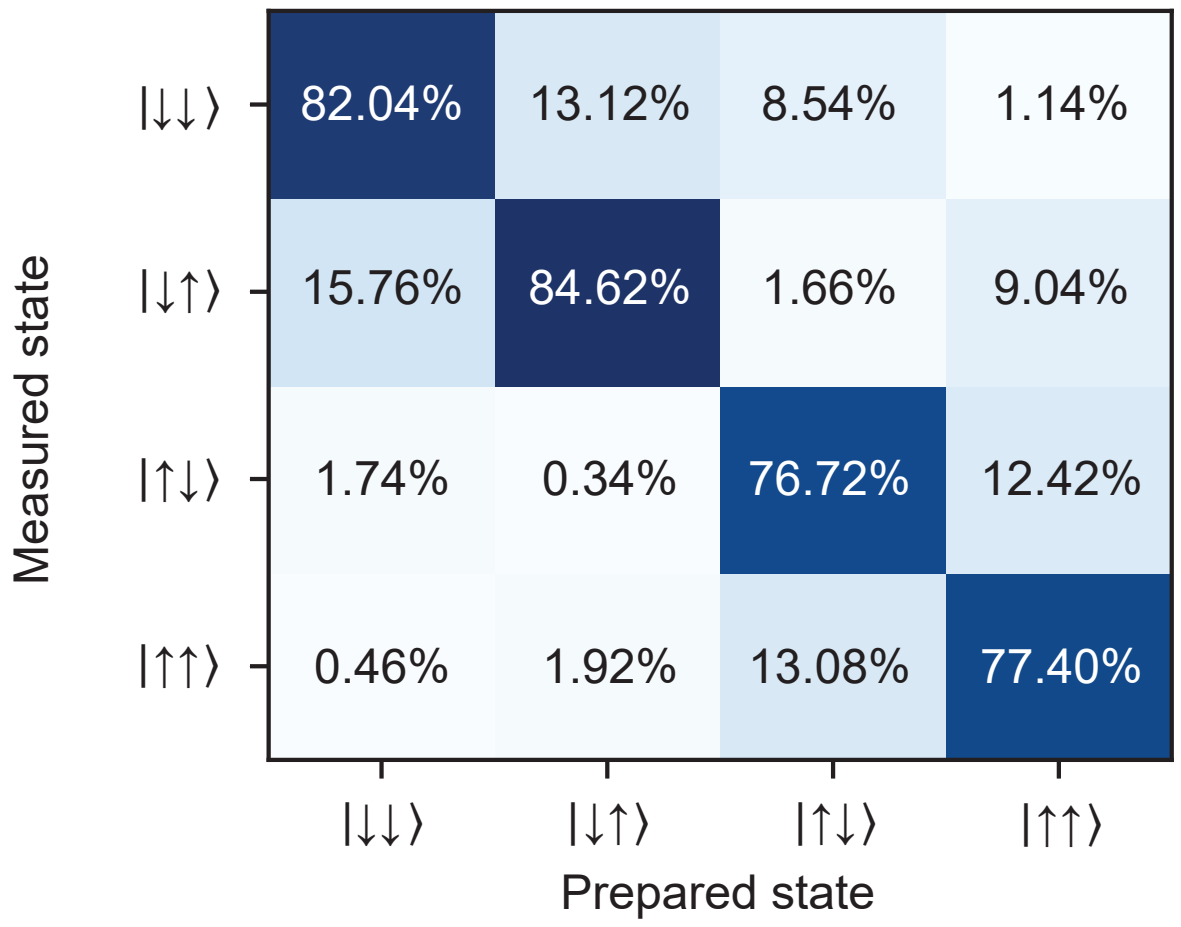

Supplementary Figure 10. **Typical SPAM matrix measured in QST experiments.**

## D. QST results

### 1. *QST of basis states and single-spin superposition states*

In order to validate the QST protocol, we first analyze the density matrices after preparing the two-qubit system in a basis state (e.g. $|\downarrow\downarrow\rangle$, $|\downarrow\uparrow\rangle$, $|\uparrow\downarrow\rangle$ or $|\uparrow\uparrow\rangle$ state). Supplementary Figure 11 displays the corresponding reconstructed density matrices with (c) and without (b) SPAM corrections. They all exhibit a single prominent component at the expected location. The bare fidelities $F$ to the target state range from 75% to 85% (see Supplementary Table 3) which provides an estimate of the overall initialization/shuttling/recovery process fidelity.

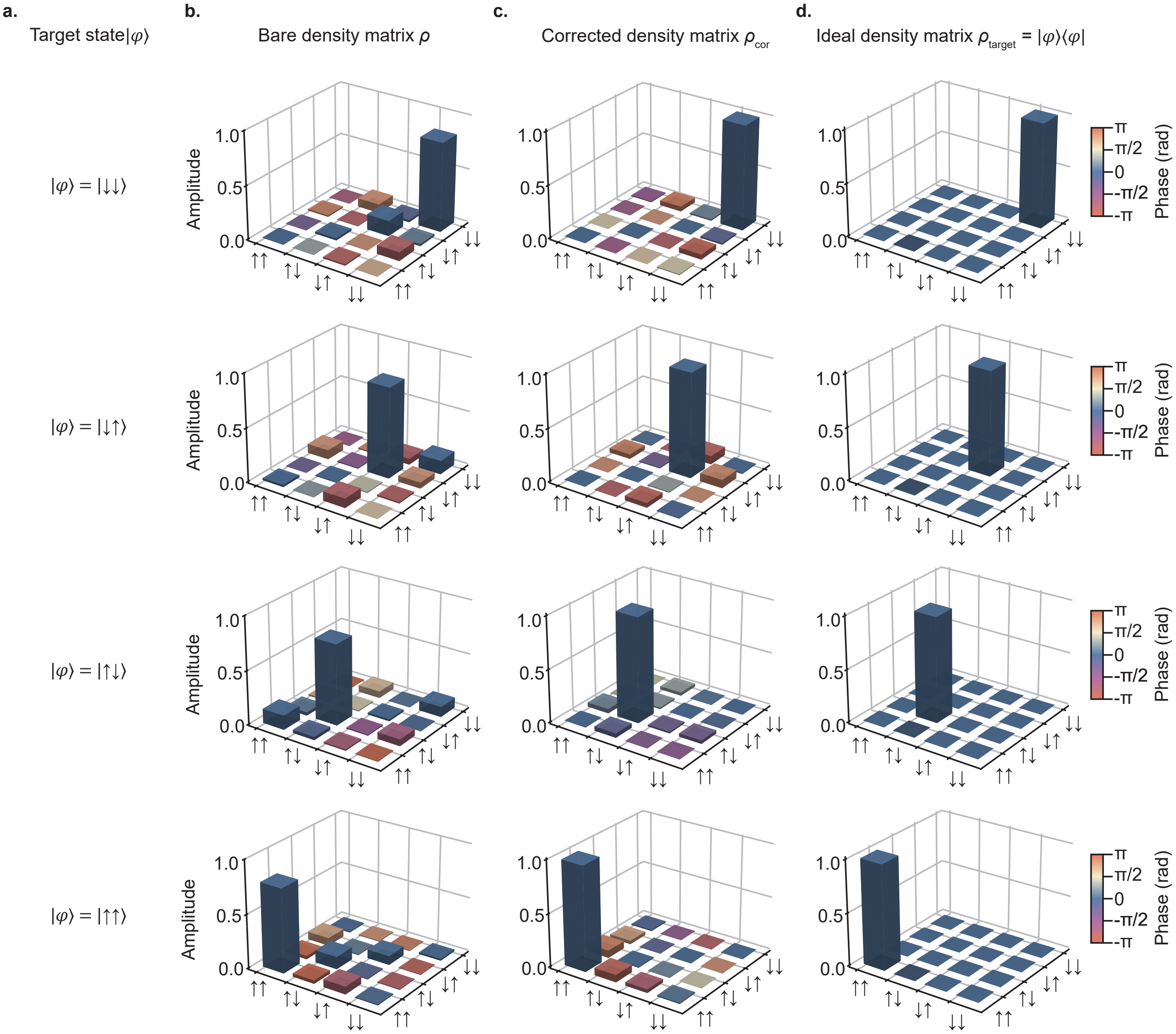

Supplementary Figure 11. **Density matrices of basis states reconstructed by QST experiments.**

When accounting for SPAM errors, the density matrices (shown in column (c)) closely match their ideal counterparts (column (d)), with fidelities exceeding 98.9%. This highlights the effectiveness of the correction protocol, but also confirms that the corrections introduced by the SPAM matrix partially compensate for shuttling and recovery gate imperfections.

Similarly, the density matrices obtained while preparing single-spin superposition states i.e. $X^{\mathrm{B}}_{\pi/2}\,|\downarrow\downarrow\rangle = \frac{1}{\sqrt{2}}\,|\downarrow\rangle \otimes (|\downarrow\rangle - i\,|\uparrow\rangle)$ and $X^{\mathrm{A}}_{\pi/2}\,|\downarrow\downarrow\rangle = \frac{1}{\sqrt{2}}(|\downarrow\rangle - i\,|\uparrow\rangle) \otimes |\downarrow\rangle$ are shown in Supplementary Figure 12. Each density matrix displays

non-diagonal elements at the expected positions with phases close to $\pm\pi/2$, in agreement with the ideal case (column (d)). The fidelities obtained $F = 85.3\%$ ($F = 73.7\%$) and $F_{\mathrm{cor}} = 99.6\%$ ($F_{\mathrm{cor}} = 97.0\%$) are similar to those reported for the basis states as expected considering the single-qubit gate fidelities reported in Supplementary Table 1. Note that here the correction by $M_{\mathrm{SPAM}}$ does not account for all errors in the calibration of the recovery gate. In particular it does not correct errors in the last rotation $Z_\gamma$.

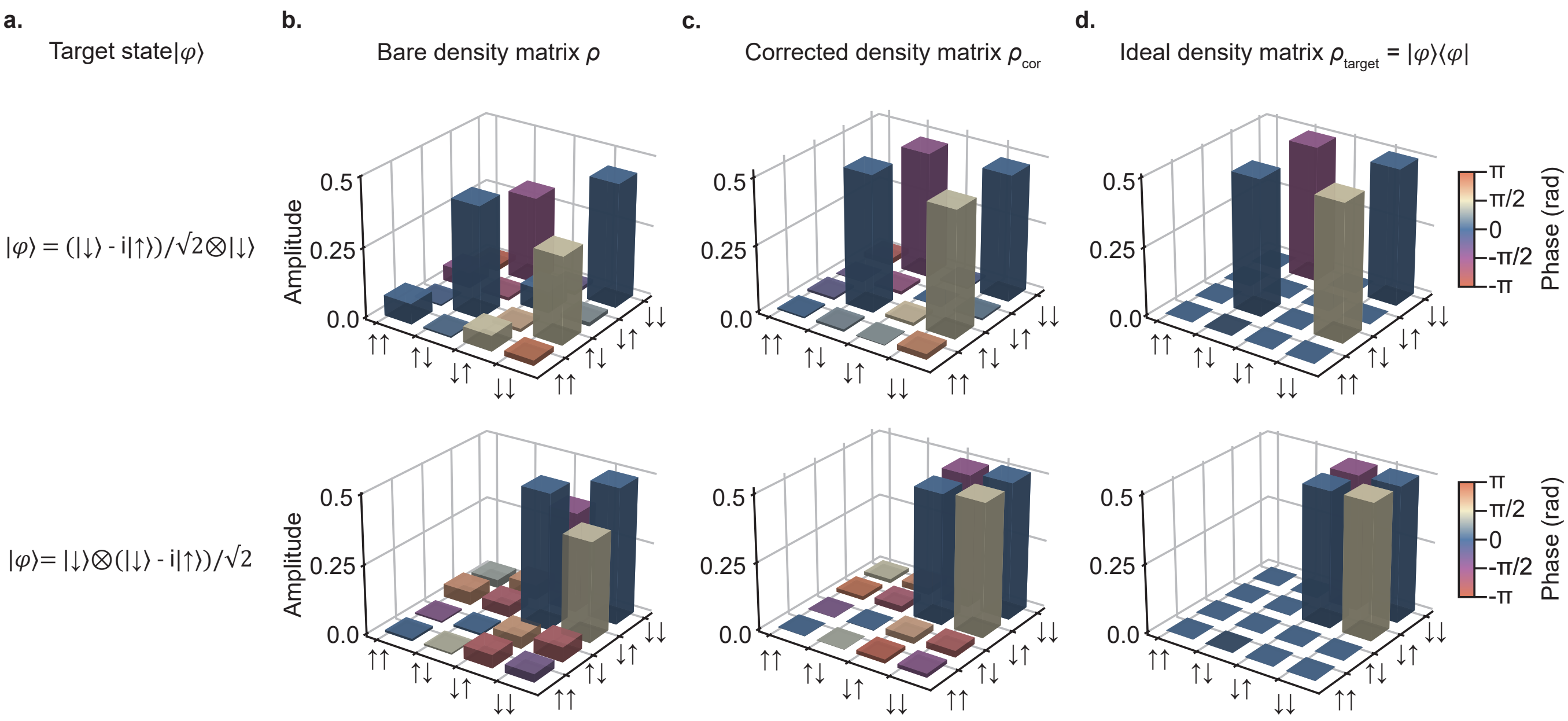

Supplementary Figure 12. **Density matrices of single-qubit superposition states reconstructed by QST.**

### 2. *QST of Bell states*

Supplementary Figure 13 shows the reconstructed density matrices for each prepared Bell state (e.g. $|\Psi^{\pm}\rangle$, $|\Phi^{\pm}\rangle$) without SPAM correction (b), with SPAM correction (c), and the corresponding ideal density matrices (d). The bare fidelities $F$ range from 61.1% to 67.1% (see also Supplementary Table 3). In all cases, they overcome the 50% threshold therefore they prove the presence of entanglement in the two-qubit system. In the case of the SPAM-corrected matrices, the estimated fidelities are lower than when preparing superposition states (Supplementary Section 8 D 1) and range from 82.5% to 92.2%. These discrepancies can be partially attributed to imperfections in the Bell state preparation and more precisely in the controlled-$Z$ gate used to entangle the qubits which usually has a lower fidelity than single-qubit operations. Additionally, the higher sensitivity of entangled states to decoherence is probably also at the origin of the discrepancy observed. The higher fidelities measured for $|\Psi^{\pm}\rangle$ states than for $|\Phi^{\pm}\rangle$ states could indicate different sensitivities to spatially-correlated noise [29].

## E. Estimation of the uncertainties on the target state fidelities using bootstrapping

Fidelity uncertainties are estimated using a bootstrapping analysis [30], in which synthetic datasets are generated by randomly resampling the complete dataset with replacement. Each of these datasets is analyzed similarly to the original data, yielding to a corresponding fidelity value. Herein, 5000 bootstrapped samples are used to evaluate the fidelity error, resulting in distributions for a bare fidelity $F$ and a corrected fidelity $F_{\mathrm{cor}}$, as shown in Supplementary Figure 14. The standard deviation ($\sigma$) of each distribution then provides a reliable estimate on the fidelity uncertainty in the original sampling. Mean values of the distribution are respectively denoted $F^{\mathrm{bst}}$ and $F^{\mathrm{bst}}_{\mathrm{cor}}$. Importantly, the corrected fidelities from synthetic datasets are computed using the previously defined SPAM matrix, that is not resampled.

Supplementary Figure 14 illustrates the bootstrapping analysis for two different target states being $|\Phi^{+}\rangle$ and $|\uparrow\uparrow\rangle$. In most cases, the fidelity distributions can be fitted with Gaussian envelopes (red and blue solid lines), allowing to identify the 95 % confidence interval as being $[F^{\text{bst}}_{(\text{correc})} - 2\sigma, F^{\text{bst}}_{(\text{correc})} + 2\sigma]$. When the mean fidelity is close to 1, as for SPAM-corrected density matrices of basis states and superposition states, the distribution is skewed and can no longer be approximated as a simple Gaussian model preventing us to determine a 95 % confidence interval. An example is shown in panel (d) of Supplementary Figure 14.

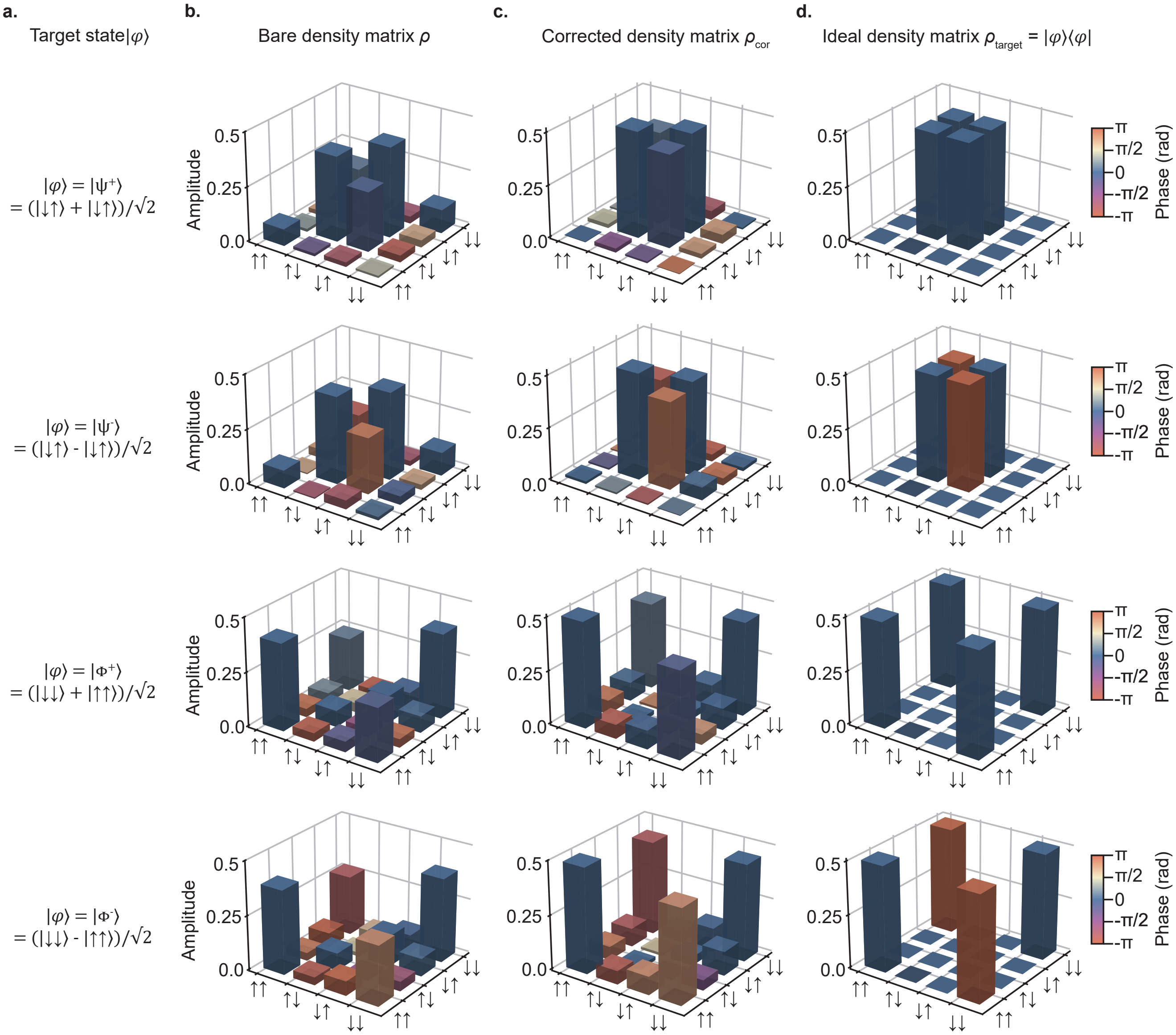

Supplementary Figure 13. **Density matrices of Bell states reconstructed by QST.**

Supplementary Table 3 summarizes all previously mentioned fidelities and highlights strong agreement between the original fidelities estimated for each target state and the values obtained from the bootstrapping analysis.

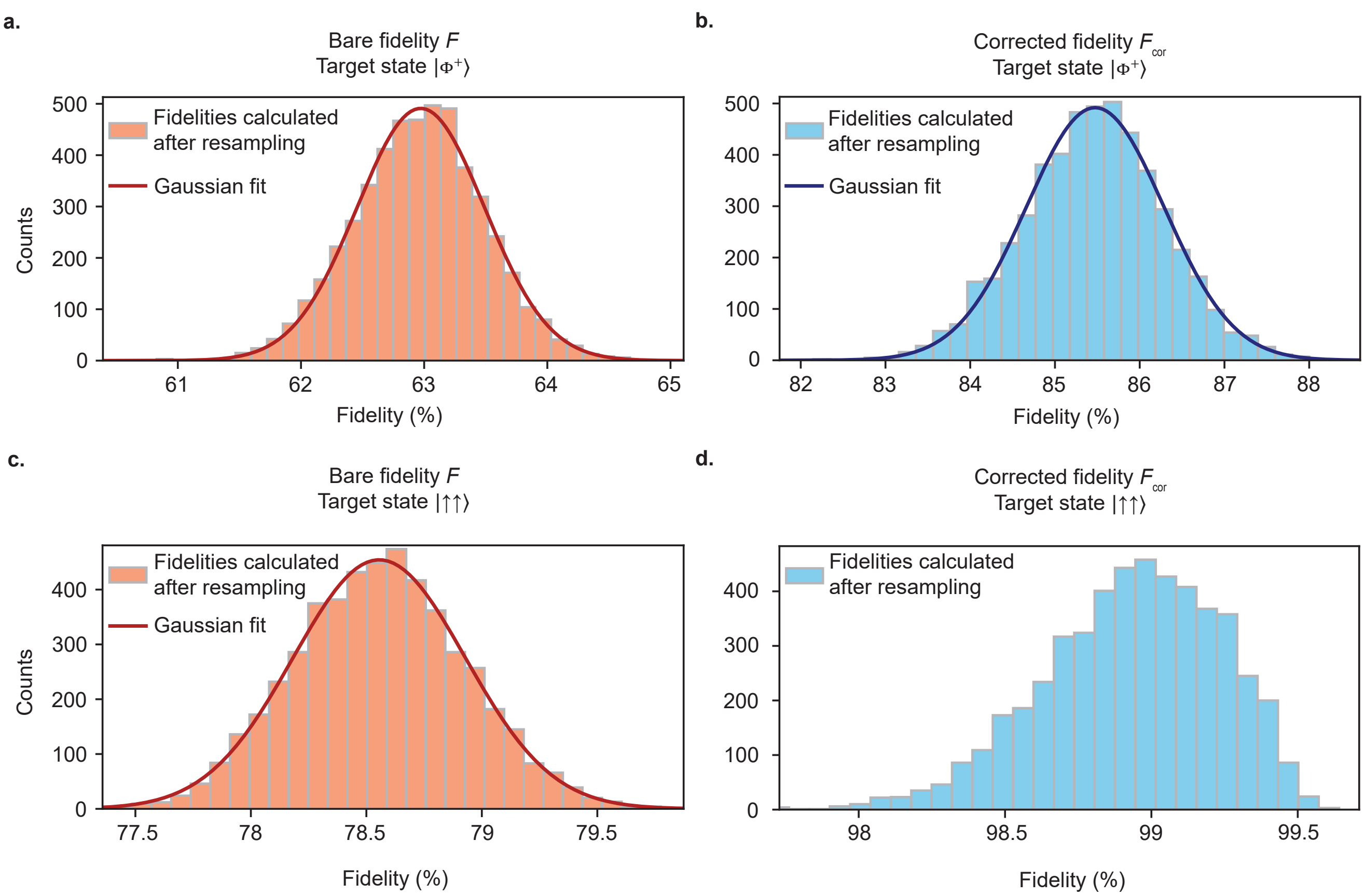

Supplementary Figure 14. **Estimation of fidelities based on bootstrapping analysis**. Results of the bootstrapping analysis for the QST experiments where **a.** and **b.** the target state is $|\Phi^+\rangle$, **c.** and **d.** the target state is $|\uparrow\uparrow\rangle$. The fidelity distributions obtained follow Gaussian distributions except in d where the correction by $M_{\mathrm{SPAM}}$ leads to fidelities very close to one and thus to a left-skewed distribution.

| Target state $\lvert\varphi\rangle$ | Bare fidelity from experiments $F$ (%) | Corrected fidelity from experiments $F_{\mathrm{cor}}$ (%) | Bare fidelity from bootstrapping $F^{\mathrm{bst}}$ (%) | Corrected fidelity from bootstrapping $F^{\mathrm{bst}}_{\mathrm{cor}}$ (%) |
|---|---|---|---|---|
| $\lvert\downarrow\downarrow\rangle$ | 83.3 | 99.7 | 83.3 $\pm$ 0.6 | 99.7 ($2\sigma = 0.1$) • |
| $\lvert\downarrow\uparrow\rangle$ | 84.7 | 98.9 | 84.7 $\pm$ 0.6 | 98.9 $\pm$ 0.3 |
| $\lvert\uparrow\downarrow\rangle$ | 75.3 | 99.7 | 75.3 $\pm$ 0.8 | 99.5 ($2\sigma = 0.5$) • |
| $\lvert\uparrow\uparrow\rangle$ | 78.6 | 99.1 | 78.6 $\pm$ 0.7 | 98.9 ($2\sigma = 0.6$) • |
| $\frac{1}{\sqrt{2}}\lvert\downarrow\rangle \otimes (\lvert\downarrow\rangle - i\lvert\uparrow\rangle)$ | 85.3 | 99.6 | 85.3 $\pm$ 0.6 | 99.6 ($2\sigma = 0.3$) • |
| $\frac{1}{\sqrt{2}}(\lvert\downarrow\rangle - i\lvert\uparrow\rangle) \otimes \lvert\downarrow\rangle$ | 73.7 | 97.0 | 73.7 $\pm$ 0.8 | 96.9 $\pm$ 0.8 |
| $\lvert\Psi^+\rangle = \frac{1}{2}(\lvert 01\rangle + \lvert 10\rangle)$ | 67.1 | 92.2 | 67.1 $\pm$ 1.0 | 92.0 $\pm$ 1.4 |
| $\lvert\Psi^-\rangle = \frac{1}{2}(\lvert 01\rangle - \lvert 10\rangle)$ | 64.7 | 88.2 | 64.7 $\pm$ 1.0 | 88.1 $\pm$ 1.6 |
| $\lvert\Phi^+\rangle = \frac{1}{2}(\lvert 00\rangle + \lvert 11\rangle)$ | 63.0 | 85.5 | 63.0 $\pm$ 1.0 | 85.5 $\pm$ 1.6 |
| $\lvert\Phi^-\rangle = \frac{1}{2}(\lvert 00\rangle - \lvert 11\rangle)$ | 61.1 | 82.5 | 61.1 $\pm$ 1.0 | 82.4 $\pm$ 1.6 |

Supplementary Table 3. **Fidelities of different target states computed by QST analysis.** The uncertainties corresponds to two standard deviations ($2\sigma$) obtained from the bootstrapping analysis. Fidelities marked by a • symbol correspond to skewed distributions that cannot be fitted with a Gaussian model. In such cases, there is no straightforward correspondence between the $\pm 2\sigma$ interval and the 95% confidence interval.